%% file: main.tex
\documentclass[11pt, letterpaper]{article}
\pdfoutput=1
  \usepackage{fullpage}

\usepackage{array,multirow,tikz,wrapfig}
\usetikzlibrary{shapes.geometric}
\usepackage{empheq}
\usepackage[most]{tcolorbox}
\newtcbox{\mymath}[1][]{%
    nobeforeafter, math upper, tcbox raise base,
    enhanced, colframe=blue!30!black,
    colback=blue!30, boxrule=1pt,
    #1}

\tikzset{base/.style={draw, align=center, minimum height=4ex},
         test1/.style={base, diamond, aspect=2, text width=5em, inner sep=5pt},
         test2/.style={base, diamond, aspect=2, text width=5em, inner sep=-4.8pt}
        }
        
\usepackage{subfig}
\usepackage{algorithm,algorithmic}
\usepackage{amsmath,amssymb,amsthm}
\usepackage{booktabs} % For formal tables
\usepackage{xspace}
\usepackage{hyperref,url}
\usepackage[papersize={8.5in,11in},top=1.5in,left=1in,right=1in,bottom=1.5in]{geometry}

\newcommand\myeq{\stackrel{\mathclap{\normalfont\mbox{def}}}{=}}
\newcommand{\var}{\sigma}

\title{Risk-Averse Matchings over Uncertain Graph Databases} 
\author{
Charalampos E. Tsourakakis, Shreyas Sekar, Johnson Lam, Liu Yang  \\
Boston University, University of Washington, Boston University, Yale University \\
\texttt{ctsourak@bu.edu,sekarshr@uw.edu, jlam17@bu.edu, liu.yang@yale.edu} \\
}

\begin{document}

\maketitle 
\input{commands}

\begin{abstract}
\input{Abstract}

\end{abstract}

\newpage

\section{Introduction}
\label{sec:introduction}
\input{Introduction.tex}

\section{Related Work}
\label{sec:related}
\input{Related.tex}

\section{Model and Proposed Method}
\label{sec:proposed}
\input{Proposed.tex}

\section{Experimental Results}
\label{sec:exp}
\input{Experiments.tex}

\section{Conclusion}
\label{sec:concl}
\input{Conclusion.tex}
\newpage 
\section*{Acknowledgements} 
Charalampos Tsourakakis would like to thank his newborn son Eftychios for the happiness he brought to his family.

\bibliography{ref}
\bibliographystyle{abbrv}

\end{document}

%% file: commands.tex
\makeatletter
\newcommand*{\rom}[1]{\expandafter\@slowromancap\romannumeral #1@}
\makeatother

\newcommand{\Oh}{\mathcal{O}}
\newcommand{\maj}{\mathrm{maj}}
\newcommand{\hide}[1]{} 
\newtheorem{claim}{Claim}
\newtheorem{lemma}{Lemma}
\newtheorem{theorem}{Theorem}
\newtheorem{definition}{Definition}
\newtheorem{corollary}{Corollary}

\newcommand{\ber}[1]{\text{Bernoulli}\left(#1\right)}
\newcommand{\bin}[2]{\text{Bin}\left(#1,#2\right)}
\newcommand{\geom}[1]{\text{Geom}\left(#1\right)}
\newcommand{\hypg}[3]{\text{Hypergeometric}\left(#1,#2,#3\right)}
\newcommand{\unif}[1]{\text{Uniform}\left(#1\right)}
\newcommand{\KL}[2]{D_{\text{KL}}({#1}||{#2} )}

\newcommand{\pr}[1]{\ensuremath{{\bf{Pr}}\left[{#1}\right]}}

\newcommand{\cc}{\text{Correlation Clustering}\xspace}
\newcommand{\cccc}{\text{2-Correlation-Clustering}\xspace}
\newcommand{\f}{\tilde{f}}
\def\e{\epsilon}
\def\hT{\widehat{T}}
 \def\r{\rho}
\newcommand{\diam}{\frac{ \log{n}}{\log{\log{n}}}}
\def\hD{\widehat{D}}
\def\g{\gamma}
\def\risk{risk}
\newcommand{\beql}[1]{\begin{equation}\label{#1}}

\newcommand{\beq}[1]{\begin{equation}\label{#1}}
\newcommand{\eeq}{\end{equation}}
\newcommand{\sfrac}[2]{\frac{\scriptstyle #1}{\scriptstyle #2}}
\newcommand{\bfrac}[2]{\left(\frac{#1}{#2}\right)}
\newcommand{\brac}[1]{\left(#1\right)}
\def\a{\alpha}
\newcommand{\misc}[1]{{\color{magenta}#1}}
\newcommand{\reminder}[1]{{\color{red}#1}}
 \newcommand{\vectornorm}[1]{\left|\left|#1\right|\right|}
\newcommand{\field}[1]{\mathbb{#1}} % requires amsfonts
\newcommand{\One}[1]{\ensuremath{{\mathbf 1}\left(#1\right)}}
\newcommand{\Prob}[1]{\ensuremath{{\bf{Pr}}\left[{#1}\right]}}
\newcommand{\Mean}[1]{\ensuremath{{\mathbb E}\left[{#1}\right]}}
\newcommand{\NP}{\ensuremath{\mathbf{NP}}\xspace}
\newcommand{\NPhard}{{\ensuremath{\mathbf{NP}}-hard}\xspace}
\newcommand{\NPcomplete}{{\ensuremath{\mathbf{NP}}-complete}\xspace}
\newcommand{\sgn}{{\ensuremath{\mathrm{sgn}}}}
\newcommand{\whp}{\textit{whp}\xspace}
\newcommand{\Var}[1]{{\mathbb Var}\left[{#1}\right]}

\newcommand{\spara}[1]{\smallskip\noindent{\bf #1}}
\newcommand{\mpara}[1]{\medskip\noindent{\bf #1}}
\newcommand{\para}[1]{\noindent{\bf #1}}

%% file: Abstract.tex
A large number of applications such as querying sensor networks,  and analyzing protein-protein interaction (PPI) networks, rely on mining  uncertain graph and hypergraph databases. In this work we study the following problem: 

\begin{quotation}
\noindent Given an uncertain, weighted (hyper)graph, how can we efficiently find a (hyper)matching with high expected reward, and low risk?
\end{quotation}

\noindent This problem  naturally arises in the context of several important applications, such as online dating, kidney exchanges, and team formation.   We introduce a novel formulation for  finding matchings with maximum expected reward and bounded risk under a general model of uncertain weighted (hyper)graphs that we introduce in this work. Our model generalizes probabilistic models used in prior work, and captures both continuous and discrete probability distributions, thus allowing to handle privacy related applications that inject appropriately distributed noise to (hyper)edge weights. Given that our optimization problem is NP-hard, we turn our attention to designing efficient approximation algorithms. For the case of uncertain weighted graphs, we provide a $\frac{1}{3}$-approximation algorithm, and a $\frac{1}{5}$-approximation algorithm with near optimal run  time.  
 For the case of uncertain weighted hypergraphs, we provide a $\Omega(\frac{1}{k})$-approximation algorithm, where $k$ is the rank of the hypergraph (i.e., any hyperedge includes at most $k$ nodes), that runs in almost (modulo log factors) linear time.

We complement our theoretical results by testing our  approximation algorithms on a wide variety of synthetic experiments, where we observe in a controlled setting interesting findings on the trade-off between reward, and risk.   We also provide an application of our formulation for providing recommendations of  teams that are likely to  collaborate, and have high impact.  Our code is available at \url{https://github.com/tsourolampis/risk-averse-graph-matchings}.

%% file: Introduction.tex
Graphs model a wide variety of  datasets that consist of a set of entities, and pairwise relations  among them. In several real-world applications, these relations  are inherently uncertain. For example, protein-protein interaction (PPI) networks are  associated with uncertainty since protein interactions are obtained via noisy, error-prone measurements \cite{asthana2004predicting}. In privacy applications deterministic edge weights become appropriately defined random variables \cite{boldi2012injecting,kearns2016private}, in dating applications each recommended link is associated with the probability that a date will be successful \cite{chen2009approximating}, in viral marketing the extent to which an idea propagates through a network depends on the `influence probability' of each social interaction~\cite{kempe2003maximizing}, in link prediction possible interactions  are assigned probabilities  \cite{liben2007link,tsourakakis2017predicting}, 
and in entity resolution  a classifier outputs for each pair of entities a probability that they refer to the same object.

Mining uncertain graphs poses significant challenges. Simple queries --such as distance queries-- on deterministic graphs become \#{\bf P}-complete (\cite{valiant1979complexity}) problems on uncertain graphs \cite{jin2011discovering}.  Furthermore, approaches that maximize the expected value of a given objective typically involve high risk solutions. On the other hand, risk-averse methods are based on obtaining several graphs samples, a procedure  that is computationally expensive, or even prohibitive for large-scale uncertain graphs.

Two remarks about uncertain graph models used in prior work that are worth making before we discuss the main focus of this work  follow. The datasets used in the majority of prior work are {\em uncertain, unweighted graphs}. There appears to be less work related to {\em uncertain, weighted hypergraphs} that are able to model a wider variety of datasets, specifically those containing more than just pairwise relationships (i.e., hyperedges).  Secondly, the  model of uncertain graphs used  in prior work \cite{bonchi2014core,huang2016truss,khan2014fast,khan2015uncertain,
kollios2013clustering,liu2012reliable,
moustafa2014subgraph,parchas2014pursuit,parchas2016uncertain,potamias2010k} are in-homogeneous random graphs \cite{bollobas2007phase}. More formally,  
let $\mathcal{G} = (V,E,p)$ be an uncertain  graph where $p: E \rightarrow (0,1]$, is the function that assigns a probability of success to each  edge independently from the other  edges.  According to the possible-world semantics \cite{bollobas2007phase,dalvi2007efficient} that interprets  $\mathcal{G} $ as a set $\{G: (V,E_G)\}_{E_G \subseteq E}$ of $2^{|E|}$ possible deterministic graphs (worlds), each defined by a subset of $E$.  The probability of observing any possible world $G(V,E_G) \in 2^{E}$ is

$$\Prob{G} = \prod\limits_{e \in E_G} p(e) \prod\limits_{e \in E\backslash E_G} (1-p(e)).$$ 

\noindent This model restricts the distribution of each edge to be a Bernoulli distribution, and does not capture various important applications such as privacy applications where  noise is injected on the weight of each edge \cite{boldi2012injecting,kearns2016private}. 

In this work, we focus on {\em risk-averse matchings over uncertain (hyper)graphs}.  To motivate our problem consider Figure~\ref{fig:toy} that shows a probabilistic graph (i.e., a 2-regular hypergraph) with two perfect matchings, $M_1 = \{ (A,B), (C,D)\}$  and $M_2 = \{ (A,C),$ $ (B,D) \}$.  Each edge $e$ follows a Bernoulli distribution with success probability $p(e)$, and is associated with a reward  $w(e)$ that is obtained only when the edge is successfully realized. These two parameters $(p(e), w(e))$   annotate each edge $e$  in Figure~\ref{fig:toy}.  The maximum weight matching {\em in expectation} is $M_1$ with expected reward $100\times \frac{1}{2} \times 2 = 100$. However, with   probability $(1-\frac{1}{2}) \times (1-\frac{1}{2}) = \frac{1}{4}$ the reward we  receive from $M_1$  equals zero. However, the second matching $M_2$  has expected reward equal to $80$ with probability 1.  In other words,  matching $M_1$ offers potentially higher reward but entails {\em higher risk} than $M_2$.  Indeed, in many situations with asymmetric rewards, one observes that high reward solutions are accompanied by higher risks and that such solutions may be shunned by agents in favor of safer options~\cite{kolata2009grant}.   

 \hide{
\begin{figure}
\centering
\begin{tikzpicture}
\node (A) at (0,2) {A};
\node (B) at (4, 2) {B};
\node (C) at (0,0) {C};
\node (D) at (4, 0) {D};
\draw [black][-][thick]  (A) edge (B) (B) edge (D) (D) edge (C) (C) edge (A);
\node at (2.1,1.8) {$(0.5, 100)$};
\node at (2.1,0.2) {$(0.5, 100)$};
\node[label=above:\rotatebox{90}{$(1,40)$}] at (-0.2,0.5) {};
\node[label=above:\rotatebox{90}{$(1,40)$}] at (4.2,0.5) {};
\draw[red,fill=red,opacity=0.2] (0,2) circle (2.2mm);
\draw[red,fill=red,opacity=0.2] (4,2) circle (2.2mm);
\draw[red,fill=red,opacity=0.2] (0,0) circle (2.2mm);
\draw[red,fill=red,opacity=0.2] (4,0) circle (2.2mm);
\end{tikzpicture}
\caption{\label{fig:toy} Probabilistic graph, each edge $e$ is annotated with $(p(e),w(e))$, its probability and its reward/weight. The matching $(A,B), (C,D)$ has higher expected weight than $(A,C), (B,D)$. However, the reward of the former matching is 0 with probability $\frac{1}{4}$, but the reward of the latter matching is 80 with probability 1. For details, see Section~\ref{sec:introduction}.
}
\end{figure}
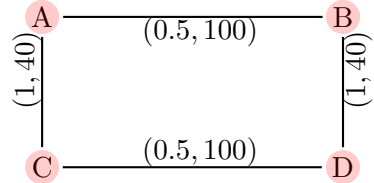
}

\begin{wrapfigure}{r}{0.4\textwidth}
\centering
\begin{tikzpicture}
\node (A) at (0,2) {A};
\node (B) at (4, 2) {B};
\node (C) at (0,0) {C};
\node (D) at (4, 0) {D};
\draw [black][-][thick]  (A) edge (B) (B) edge (D) (D) edge (C) (C) edge (A);
\node at (2.1,1.8) {$(0.5, 100)$};
\node at (2.1,0.2) {$(0.5, 100)$};
\node[label=above:\rotatebox{90}{$(1,40)$}] at (-0.2,0.5) {};
\node[label=above:\rotatebox{90}{$(1,40)$}] at (4.2,0.5) {};
\draw[red,fill=red,opacity=0.2] (0,2) circle (2.2mm);
\draw[red,fill=red,opacity=0.2] (4,2) circle (2.2mm);
\draw[red,fill=red,opacity=0.2] (0,0) circle (2.2mm);
\draw[red,fill=red,opacity=0.2] (4,0) circle (2.2mm);
\end{tikzpicture}
\caption{\label{fig:toy} Probabilistic graph, each edge $e$ is annotated with $(p(e),w(e))$, its probability and its reward/weight. The matching $(A,B), (C,D)$ has higher expected weight than $(A,C), (B,D)$. However, the reward of the former matching is 0 with probability $\frac{1}{4}$, but the reward of the latter matching is 80 with probability 1. For details, see Section~\ref{sec:introduction}.}
\end{wrapfigure}

Another way to observe that matching $M_1$ entails greater risk is to draw graph samples from this probabilistic graph multiple times, and observe that around 25\% of the realizations of $M_1$ result in zero reward. However, sampling is computationally expensive on large-scale uncertain graphs. Furthermore,  in order to obtain statistical guarantees, a large number of samples may be needed \cite{parchas2014pursuit} which makes the approach   computationally intensive or  infeasible even for medium-scale graphs. Finally, it is challenging and sometimes not always clear how to aggregate different samples \cite{parchas2014pursuit}. These two drawbacks are well-known to the database community, and recently Parchas et al.~\cite{parchas2014pursuit} suggested a heuristic to extract representative instances of uncertain graphs. While their work makes an important practical contribution, their method  is an intuitive heuristic whose theoretical guarantees and worst-case running time are not well understood~\cite{parchas2014pursuit}.  

Motivated by these concerns, we focus on the following central question: 

\begin{quotation}
How can we design {\em efficient}, {\em risk-averse algorithms} with {\em solid  theoretical guarantees} for finding maximum weight matchings in uncertain weighted graphs and hypergraphs? 
\end{quotation}

This question is well-motivated, as it naturally arises in several important applications.  In online dating applications a classifier may output a probability distribution for the probability of matching two humans successfully~\cite{tu2014online}. In kidney exchange markets,  a kidney exchange is successful according to some probability distribution that is determined by a series of medical tests. Typically, this distribution is unknown but its parameters  such as the mean and the variance can be  empirically estimated \cite{chen2009approximating}.  Finally, the success of any large organization that employs skilled human resources crucially depends  on the choice of teams that will work on its various projects. Basic team formation algorithms output a set of teams (i.e., hyperedges) that combine a certain set of desired skills  \cite{anagnostopoulos2012online,gajewar2012multi,kargar2012efficient,kargar2011discovering,
lappas2009finding,majumder2012capacitated}.   A classifier can leverage features 
that relate to crowd psychology, conformity, group-decision making,  valued diversity, mutual trust, effective and participative leadership \cite{katzenbach2000peak} to estimate the probability of success of a team.

 In detail, our contributions are summarized as follows.

\spara{Novel Model and Formulation.} We propose a general model for weighted uncertain (hyper)graphs, and  a  novel formulation 
for risk-averse maximum matchings.  Our goal is to select (hyper)edges that have \emph{high expected reward, but also bounded risk of failure}. Our problem is a novel variation of the well-studied stochastic matching problem \cite{bansal2012lp,chen2009approximating}.

\spara{Approximation algorithms.} We design efficient approximation algorithms.  For the case of uncertain graphs, using Edmond's blossom algorithm~\cite{edmonds1965paths} as a black-box, we provide a risk-averse solution that is  a $\frac{1}{3}$-approximation of the optimal risk-averse solution. Similarly, using a greedy matching algorithm as a black box we obtain a  $\frac{1}{5}$-risk-averse approximation. For hypergraphs of rank $k$ (i.e., any hyperedge contains at most $k$ nodes) we obtain a risk-averse  $\Omega(\frac{1}{k})$-approximation guarantee.  Our algorithms are risk-averse, do  not need to draw graph samples, and come with solid theoretical guarantees. Perhaps more importantly, the proposed algorithms that are based on greedy matchings have a running time of $O(m\log^2 m + n \log m)$, where $n,m$ represent the number of nodes, and (hyper)edges in the uncertain (hyper)graph respectively- this makes the algorithm easy to deploy on large-scale real-world networks such as the one considered in our experiments (see Section~\ref{sec:exp}).  

 \spara{Experimental evaluation.} We evaluate our proposed algorithm on a wide variety of synthetic experiments,  where we observe interesting findings on the trade-offs between reward and risk.   There appears to be little (or even no) empirical work on {\em uncertain, weighted hypergraphs}.   We use the Digital Bibliography and Library Project (DBLP) dataset to create a hypergraph where each node is an author, each hyperedge is a team of co-authors for each paper, the probability of a hyperedge is the probability of collaboration estimated from historical data, and the weight of a hyperedge is its citation count. This uncertain  hypergraph is particularly interesting as there exist edges with high reward (citations) but whose authors have low probability to collaborate.  On the other hand, there exist papers with a decent number of citations whose co-authors consistently collaborate. Intuitively, the more risk-averse we are, the more we should prefer the latter hyperedges.  We evaluate our proposed method on this real dataset, where we observe several interesting findings. The code and the datasets will become publicly available at \url{https://github.com/tsourolampis/risk-averse-graph-matchings}.

%% file: Related.tex
\spara{Uncertain graphs.}  Uncertain graphs naturally model various datasets including protein-protein interactions   \cite{asthana2004predicting,krogan2006global}, kidney exchanges \cite{roth2004kidney}, dating applications \cite{chen2009approximating}, sensor networks whose  connectivity links are uncertain due to various kinds of failures \cite{saha2004modeling},   entity resolution   \cite{moustafa2014subgraph},  viral marketing \cite{kempe2003maximizing}, and privacy-applications \cite{boldi2012injecting}.

Given the increasing number of applications that involve uncertain graphs,  researchers have put a lot of effort in developing algorithmic tools that tackle several important graph mining problems, see \cite{bonchi2014core,huang2016truss,khan2014fast,khan2015uncertain,
kollios2013clustering,liu2012reliable,
moustafa2014subgraph,parchas2014pursuit,parchas2016uncertain,potamias2010k}.
However, with a few exceptions these methods suffer from a critical drawback; either they are not risk-averse, or they rely on obtaining many graphs samples. Risk-aversion has been implicitly discussed  by Lin  et al.  in their work on reliable clustering \cite{liu2012reliable}, where the authors show that interpreting probabilities as weights does not result in good clusterings.   Jin et al. provide a risk-averse algorithm for distance queries on uncertain graphs \cite{jin2011discovering}. 
Parchas et al.  have proposed a heuristic to extract a good possible world in order to combine risk-aversion with efficiency  \cite{parchas2014pursuit}. However,  their work comes with no guarantees. 

\spara{Graph matching} is a major topic in combinatorial optimization. The interested reader should confer the works of Lov{\'a}sz and Plummer \cite{lovasz2009matching} for a solid exposition. Finding maximum matchings in weighted graphs  is solvable in polynomial time~\cite{edmonds1965paths,gabow90}. A faster algorithm sorts the edges by decreasing weight, and adds them to a matching greedily. This algorithm is a $\frac{1}{2}$-approximation to the optimum matching.   Finding a maximum weight hypergraph matching is NP-hard, even in unweighted 3-uniform hypergraphs (aka 3-dimensional matching) \cite{karp1972reducibility}. The greedy algorithm provides a $\frac{1}{k}$-approximation (intuitively for each hyperedge we greedily add to the matching, we lose at most $k$ hyperedges) where $k$ is the maximum cardinality of an edge.  

%Finally, we also note that other works on budgeted matchings on graphs and hypergraphs strictly generalize the problem studied in this work~\cite{cyganGM13,bergerBGS11}. However, their reliance on linear programming based techniques make them impractical for large-scale networks.  

\spara{Stochastic Matchings.} Various stochastic versions of graph matchings have been studied in the literature. We discuss two papers that lie close to our work \cite{bansal2012lp,chen2009approximating}. Both of these works consider a random graph model with a Bernoulli distribution on each edge, i.e., a graph $G([n],E)$ on $n$ nodes, where each edge $(i,j) \in {[n] \choose 2}$  exists with probability $p_{ij}$, independent of other edges. In contrast to our work, these models allow the central designer to \emph{probe} each edge to verify its realization: if the each edge exists, it gets irrevocably added to the matching. While Chen et al.~\cite{chen2009approximating} provide a constant factor approximation on unweighted graphs based on a simple greedy approach, Bansal et al.~\cite{bansal2012lp} obtain a $O(1)$-factor for even weighted graphs using an LP-rounding algorithm. On the other hand, our work focuses on designing fast algorithms that achieve good matchings with bounded risk on weighted graphs without probing the edges. Finally, since the hypergraph matching problem is also known as the set packing problem, the above problems are special cases of  stochastic set packing \cite{dean2005adaptivity}.

\spara{Risk-averse optimization} is a major topic in operations research, control theory, and finance. The typical setting of risk-averse optimization is the following: suppose that $f(\omega,X)$ is a cost function of a random variable $X$, and a decision variable $\omega$.  Different choices of $\omega$ lead to different values of the mean  $\Mean{ f(\omega,X)}$. 
There is also a {\em risk} function $R(f(\omega,X))$ associated with $f$. 
The goal of risk-averse optimization  is to choose $\omega$ such that both $\Mean{f}$ and $R(f)$ are small.  This framework captures optimization problems that arise in a number of environments with uncertainty. For example, modern portfolio theories of investment are based on the idea that risk-averse investors 
should maximize expected profit conditional on a given level of market risk. This is intuitive as higher rewards come with higher risk \cite{markowitz1991foundations} in markets.   Other examples of risk-averse optimization include risk averse \cite{ruszczynski2010risk}, 
risk averse stochastic shortest paths \cite{bell2009hyperstar}, risk averse linear/quadratic/Gaussian control \cite{whittle1981risk}, 
risk averse  covering of integer programs
\cite{srinivasan2007approximation,swamy11},  and risk averse bandit arm selection \cite{yu2013sample}. %For more details, the reader is asked to refer to the lecture notes by Shapiro and Dentcheva \cite{shapiro2009lectures}.  

%% file: Proposed.tex
\spara{Uncertain Weighted Bernoulli hypergraphs.}  Before we define a general model for uncertain weighted hypergraphs that allows for both continuous and discrete probability distributions, we  introduce a simple probabilistic model for weighted uncertain hypergraphs that generalizes the existing model for random graphs. Each edge $e$ is distributed as a weighted Bernoulli variable independently from the rest: with probability $p(e)$ it exists, and its weight/reward is equal to $w(e)$, and with the remaining probability $1-p(e)$ it does not exist, i.e., its weight is zero.   More formally, let $\mathcal{H} = ([n],E,p,w)$ be an uncertain hypergraph on $n$ nodes with $|E|=m$ potential hyperedges, where $p: E \rightarrow (0,1]$, is the function that assigns a probability of existence to each hyperedge independently from the other hyperedges, and $w:E \rightarrow \field{R}^+$. The value $w(e)$ is the reward we receive from  hyperedge $e$ if it exists. Let $r_e \myeq p(e)w(e)$ be the expected reward from edge $e$. According to the possible-world semantics \cite{bollobas2007phase,dalvi2007efficient},   the probability of observing any possible world $H(V,E_H) \in 2^{E}$  where each hyperedge $e \in E_H$ has weight $w(e)$ is 

$$\Prob{H} = \prod\limits_{e \in E_H} p(e) \prod\limits_{e \notin E_H} (1-p(e)).$$

\spara{Uncertain Weighted hypergraphs.}   More generally, let  $\mathcal{H}([n],E, \{ f_e(\theta_e) \}_{e \in E})$ be an uncertain  hypergraph on $n$ nodes,  with hyperedge set $E$. The reward $w(e)$ of each hyperedge $e \in E$ is drawn according to some probability distribution $f_e$ with parameters $\vec{\theta_e}$, i.e., $w(e) \sim f_e(x;\vec{\theta_e})$. We assume that the reward for each hyperedge is drawn independently from the rest;  each probability distribution is assumed to have finite mean, and finite variance. Given this model, we define the probability of a given hypergraph $H$ with weights $w(e)$ on the hyperedges as:
$$\Prob{H;\{ w(e)\}_{e \in E}} = \prod_{e \in E} f_e(w(e); \vec{\theta_e}).$$ 
% this assumption is not restrictive in the context of our problem; an edge with infinite variance can be safely discarded. 
 
For example, suppose the reward $w(e)$ of  hyperedge $e$ is distributed as a normal random variable $\mathcal{N}(r_e, \sigma_e^2)$. Then, the probability of a hypergraph $H$ is 

$$ \Prob{H;\{ w(e)\}_{e \in E}} =\prod_{e \in E} \frac{1}{\sqrt{2\pi} \sigma_e} e^{ -\frac{(w(e)-r_e)^2}{2\sigma_e^2} }.$$ 

\noindent Our model allows for both discrete and continuous distributions, as well as mixed discrete and continuous distributions. In our experiments (Section~\ref{sec:exp}) we focus on the weighted Bernoulli, and Gaussian cases.
 
\spara{Problem definition.}   In contrast to prior work on stochastic  matchings \cite{bansal2012lp,chen2009approximating}, we do not probe edges to verify their existence; our goal is  to output a matching $M$ with high expected reward and low variance. Formally, let $\mathcal{M}$ be the set of all matchings from the hyperedge set $E$. The total associated reward with a matching $M \in \mathcal{M}$ is the expected reward, i.e., 

$$ R(M) \myeq  \sum\limits_{e \in M} r_e =  \sum\limits_{e \in M}  p(e)w(e).$$ 

\noindent Similarly, the associated risk in terms of the standard deviation  is defined as 

$$ risk(M)   \myeq  \sum\limits_{e \in M} \var_e,$$ 

\noindent  where $\var_e$ denotes the standard deviation of the distribution $f_e(x;\vec{\theta_e})$. 
 
 Given an uncertain weighted hypergraph, and a risk upper-bound $B$, our goal is to maximize the expected reward over all matchings with risk at most $B$. We refer to this problem as the  {Bounded Risk Maximum Weighted Matching} (BR-MWM)  problem. Specifically,
 
\begin{tcolorbox}
\begin{align}
\label{eq:opmin1}
\begin{array}{ll@{}ll}
\max\limits_{  M \in \mathcal{M}}  &  R(M)  &[ \text{BR-MWM~ problem}]  &   \\
\text{s.t} &   risk( M )  \leq B & & \\ 
\end{array}
\end{align}
\end{tcolorbox}

In the case  of uncertain weighted Bernoulli hypergraphs, Formulation~\eqref{eq:opmin1} becomes 

\begin{align}
\label{eq:bernoulli}
\begin{array}{ll@{}ll}
\max\limits_{M \in \mathcal{M}}  &  \sum_{e \in M} p(e)w(e)  &  &   \\
\text{s.t} &    \sum\limits_{e \in M}  w_e \sqrt{ p(e)(1-p(e))} \leq B & & \\ 
\end{array}
\end{align}  

\noindent and in the case of uncertain weighted Gaussian hypergraphs

\begin{align}
\label{eq:gaussian}
\begin{array}{ll@{}ll}
\max\limits_{M \in \mathcal{M}}  &  \sum_{e \in M}r_e  &  &   \\
\text{s.t} &    \sum\limits_{e \in M}  \var_e \leq B. & & \\ 
\end{array}
\end{align}

Finally, we remark that the BR-MWM problem is NP-Hard even on graphs via a simple reduction from Knapsack.

\spara{Other Measures of Risk.} It is worth outlining that our model and proposed method adapts easily to other risk measures.  For example, if we define the risk of a matching $M$ in terms of its variance, i.e.,

\begin{equation}
 risk(M)   \myeq   \sum\limits_{e \in M} \var_e^2,
\label{eqn_risk_variance}
\end{equation}

\noindent then all of our theoretical guarantees and the insights gained via our experiments still hold with minor changes in the algorithm. At the end of this section, we discuss in detail the required changes. For the sake of convenience and concreteness, we present our results in terms of the $\ell_1$ version of the risk (the standard deviation).

\spara{An LP-approximation algorithm.} The  {\em Hypermatching Assignment Problem} (HAP) was introduced by Cygan et al. \cite{cyganGM13}: given a $k$-uniform hypergraph $H(V,E)$, and a set of $q$ clients, each with a budget $B_i \geq 0, i=1,\ldots,q$,  a profit and a cost $w_{i,e}, b_{i,e} \geq 0$ for hyperedge $e$ respectively, the goal is to compute a matching $M$, and partition $M$ into $q$ subsets $M_1,\ldots,M_q$ so that the total profit $\sum_{i=1}^q w_i(M_i) = \sum_{i=1}^q \sum_{e \in M_i} w_{i,e}$  is maximized and the budget constraint $\sum_{e \in M_i} b_{i,e} \leq B_i$ is satisfied for all clients $i$. Our  BR-MWM  problem is a special case of HAP where there is one client ($q=1$), the profit $w_{1,e}$ is the expected reward $r_e$, and the cost $b_{1,e}$ is the standard deviation $\var_e$. Notice that without any loss of generality we can convert the uncertain hypergraph $H$ to a $k$-uniform hypergraph where $k$ is the maximum cardinality of a hyperedge by adding dummy nodes.  Therefore, we can invoke the randomized $\frac{1}{k+1+\epsilon}$-approximation algorithm for HAP \cite{cyganGM13} to solve our problem, here $\epsilon > 0$ is constant. However,  this approach --at least for the moment-- is unlikely to scale  well:  it requires solving a linear program with an exponential number of variables in terms of $\frac{1}{\epsilon}$, and then strengthen this LP by one round of the Lasserre's lift-and-project method. This motivates the design of scalable approximation algorithms.

\spara{Algorithm.}  Our algorithm is described in pseudocode~\ref{pseudo_alg_matchings}. It takes as input an uncertain weighted hypergraph as well as a hypergraph matching algorithm $MATCH$-$ALG$ as a black-box: the black-box takes a weighted hypergraph and returns a hypergraph matching. First, our algorithm removes all hyperedges that have negative reward as they are not part of any optimal solution. Similarly, it removes any edge $e$ for which $\var_e > B$; since the risk of any matching is the sum of the standard deviation of its edges, any such edge cannot be part of any optimal solution either. For any given edge $e \in E$, define $\alpha_e \myeq \frac{r_e}{\var_e}$.  Now,  we label the edges in $E$ as $e_1, e_2, \ldots, e_{m}$ such that $\alpha_{e_1} \geq \alpha_{e_2} \geq \ldots \geq \alpha_{e_{m}}$, breaking ties arbitrarily.   Sorting the $\alpha$ values requires $O(m\log m)$ time.   Next, we consider the nested sequence of hypergraphs $\emptyset = H^{(0)} \subset H^{(1)} \subset \ldots \subset H^{(m)}=H$, where $H^{(i)}$ contains the $i$ hyperedges $(e_1, e_2, \ldots, e_i)$, and each edge $e$ is weighted by the expected reward $r_e$. 

Let $M^{(i)}$ be the matching returned by {\sc Match-Alg} on $H^{(i)}$ with weights $(r_e)_{e \in H^{(i)}}$. We first compute the maximum weight matching on $H^{(m)}$. If the quantity $risk(M^{(m)})$ is less than or equal to $B$, then we output $M^{(m)}$. Otherwise, we binary search the nested sequence of hypergraphs to find {\em any} index $\ell^*$ for which 
$$risk(M^{(\ell^*)}) \leq B < risk(M^{(\ell^*+1)}).$$

\begin{wrapfigure}{l}{0.4\textwidth}
\centering
\begin{tikzpicture}
\node (A) at (0,2) {A};
\node (B) at (4, 2) {B};
\node (C) at (0,0) {C};
\node (D) at (4, 0) {D};
\draw [black][-][thick]  (A) edge (B) (B) edge (D) (D) edge (C) (C) edge (A);
\node at (2.1,1.8) {$(1.5,0.5,3)$};
\node at (2.1,0.2) {$(0.1,1,0.1)$};
\node[label=above:\rotatebox{90}{$(1,0.1,10)$}] at (-0.4,0.11) {};
\node[label=above:\rotatebox{90}{$(1,0.35,2.9)$}] at (4.5,0.04) {};
\draw[red,fill=red,opacity=0.2] (0,2) circle (2.2mm);
\draw[red,fill=red,opacity=0.2] (4,2) circle (2.2mm);
\draw[red,fill=red,opacity=0.2] (0,0) circle (2.2mm);
\draw[red,fill=red,opacity=0.2] (4,0) circle (2.2mm);
\end{tikzpicture}
\caption{\label{fig:counterexample} The risk $risk(M^{(i)})$ of the optimum matching $M^{(i)}$ is \emph{not} monotonically increasing with $i$. For details, see Section~\ref{sec:proposed}.}
\end{wrapfigure}
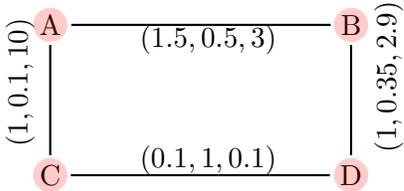 

 The final output matching $M_{OUT}$ is either $M^{(\ell^*)}$ or $e_{\ell^*+1}$, depending on which one achieves greater expected reward.  Intuitively, the latter case is required when there exists a single high-reward hyperedge whose risk is comparable to the upper bound $B$. In general, there may be more than one index that satisfies the above condition since  the variance is \emph{not} monotonically increasing with $i$. Figure~\ref{fig:counterexample} provides such an example that shows that increasing the set of allowed edges can actually decrease the overall risk of the optimum matching. 
 Specifically,  Figure~\ref{fig:counterexample} shows an uncertain graph, each edge $e$ is annotated with $(r_e,\var_e,\alpha_e)$. One can always find distributions that satisfy these parameters. We consider Algorithm~\ref{pseudo_alg_matchings} with the black-box matching algorithm {\sc Match-Alg} as the optimum matching algorithm on weighted graphs. As our algorithm considers edges in decreasing order of their $\alpha$-value, we get that $M^{(1)} = \{(A,C)\}, M^{(2)} = \{(A,B)\}, M^{(2)} = \{(A,B)\}, M^{(3)} = \{(A,C), (B,D)\}$. The risk of the above three matchings are $0.1, 0.5$, and $0.45$ respectively. Thus, the risk $risk(M^{(i)})$ of the optimum matching $M^{(i)}$ is \emph{not} monotonically increasing with $i$.

  \begin{algorithm*}[!ht]
\begin{algorithmic}  
\REQUIRE  $\mathcal{H}([n],E, \{ f_e(\theta_e) \}_{e \in E})$, Black-box algorithm {\sc Match-Alg}
\STATE Let $r_e,\var_e$ be the expectation, and the standard deviation (s.t.d) of $f_e$ for each hyperedge $e \in E$
\STATE Remove all hyperedges that have either non-positive reward ($r_e\leq 0$, or s.t.d greater than $B$ ($\var_e>B$) 
\COMMENT{Such edges are not part of any optimal solution.}
\STATE Sort the hyperedges in decreasing order according to $\alpha_e = \frac{ r_e }{\var_e}$, let $\alpha_{e_1} \geq \ldots \geq \alpha_{e_m} \geq 0$.
\STATE $M^{(m)}\leftarrow $ {\sc Match-Alg}($H^{(m)}$)
\IF{ $risk(M^{(m)}) \leq B$ }
\STATE $\ell^* \leftarrow m$ 
\STATE Return $\ell^*, M^{(\ell^*)}$
\ENDIF
\STATE $low \leftarrow 1, high \leftarrow m$ 
\WHILE{True}
 \STATE $mid \leftarrow \lfloor \frac{low+high}{2} \rfloor$
 \STATE Compute $M^{(mid)}, M^{(mid+1)}$
\IF{$risk(M^{(mid)}) \leq B < risk(M^{(mid+1)})$}
\STATE $\ell^* \leftarrow mid$
\STATE Return $\ell^*, M^{(\ell^*)}$
\ELSIF{ $\risk(M^{(mid)}) \leq B$} 
\STATE  $low \leftarrow mid+1$ 
\ELSE 
\STATE $high \leftarrow mid$ 
\ENDIF 
\ENDWHILE
\end{algorithmic}
\caption{\label{pseudo_alg_matchings} Algorithm for computing a $\frac{c}{2+c}$-approximate matching for the BR-MWM problem  on uncertain weighted hypergraphs.}	
\end{algorithm*}

While it is not hard to see how a binary search would work, we provide the details for completeness. We know that $risk(M^{(1)})=\var(e_1) \leq B$, and $risk(M^{(m)}) > B$. Let $low=1,high=m$. We search the middle position $mid$ between low and high, and $mid+1$. If $risk(M^{(mid)}) \leq B < risk(M^{(mid +1)})$, then we set $\ell^*$ equal to $mid$ and return. If not, then if $risk(M^{(mid)}) \leq B$, we repeat the same procedure with $low = mid+1, high = m$. Otherwise, if 
$risk(M^{(mid)}) > B$ we repeat   with $low =1, high=mid$. This requires $O(\log m)$ iterations, and each iteration requires at most two maximum weighted matching computations.  

Our proposed algorithm uses the notion of a black-box reduction: wherein, we take an arbitrary $c$-approximation algorithm for computing a maximum-weight hypermatching ({\sc Match-Alg}, $c\leq 1$) and leverage its properties to derive an algorithm that in addition to maximizing the expected weight also has low risk. This black-box approach has a significant side-effect: organizations may have already invested in graph processing software for deterministic graphs, 
that they would like to use regardless of the uncertainty inherent in the data. Our search takes  time $O(\log m\times T(n,m) )$ where $T(n,m)$ is the running time of maximum weighted matching algorithm {\sc Match-Alg}.

\spara{$\frac{1}{3}$-approximation} \spara{for uncertain weighted graphs.}  First we analyze our algorithm for the important case of uncertain weighted graphs. Unlike general hypergraphs, we can find a maximum weight graph matching in polynomial time  using Edmond's algorithm \cite{gabow90}. Our main result is stated as the following theorem.

\begin{theorem}
\label{thm_graphmatchings}
Assuming an exact maximum weight matching algorithm {\sc Match-Alg}, Algorithm~\ref{pseudo_alg_matchings}  returns a matching $M_{OUT}$ whose risk is less or equal than $B$, and whose expected reward is at least $\frac{1}{3}$ of the optimal solution to the Bounded Risk Maximum Weighted Matching problem on uncertain weighted graphs.
\end{theorem}

Before we prove Theorem~\ref{thm_graphmatchings}, it is worth pointing out, that besides the fact that our proposed algorithm  can be easily implemented using existing graph matching software,  it also provides a better approximation  than the approximation achieved using  \cite{cyganGM13}, i.e., $\frac{1}{3}>\frac{1}{3+\epsilon}$ for any constant $\epsilon>0$.

\begin{proof}
Let $M^{OPT}$ denote an optimum matching whose risk is at most $B$. Since it is immediately clear by the description of our algorithm that $risk(M_{OUT}) \leq B$, our goal is to prove that the matching returned by our algorithm has reward at least one-third as good as the reward of the optimum matching, i.e., $R(M_{OUT})= \sum\limits_{e \in M_{OUT}} r_e  \geq \frac{R(M^{OPT})}{3}$. 
 
 \noindent In order to prove this bound, we prove a series of inequalities. By definition, $H^{(\ell^*+1)}$ differs from $H^{(\ell^*)}$ in exactly one edge, that is $e_{\ell^*+1}$. We also know that the maximum weight matching in $H^{(\ell^*+1)}$ is different from the maximum weight matching in $H^{(\ell^*)}$ since the former entails risk that exceeds the  budget $B$. We conclude that $M^{(\ell^*+1)}$ contains the edge $e_{\ell^*+1}$. 

Therefore, we have that $ R(M^{(\ell^*+1)}) = R(M^{(\ell^*+1)} \setminus e_{\ell^*+1})  + r(e_{\ell^*+1}) \leq R(M^{(\ell^*)})  + r(e_{\ell^*+1})$. This is true because  $M^{(\ell^*)}$ is the maximum weight matching in $H^{(\ell^*)}$ and so its weight is larger than or equal to that of $M^{(\ell^*+1)} \setminus e_{\ell^*+1}$. In conclusion, our first non-trivial inequality is:
\begin{equation}
\label{eqn_ml1}
R(M^{(\ell^*)}) +  r(e_{\ell^*+1}) \geq R(M^{(\ell^*+1)})
\end{equation}

\noindent Next, we   lower-bound  $M^{(\ell^*+1)}$ by using the facts that $\alpha_e \geq \alpha_{e_{\ell^*+1}}$ for all $e \in M^{(\ell^*+1)}$, and  that the total risk of $M^{(\ell^*+1)}$ is at least $B$ by definition. Specifically, 
\allowdisplaybreaks
\begin{align}
 R(M^{(\ell^*+1)})  & = \sum_{e \in M^{(\ell^*+1)}}r_e   = \sum_{e \in M^{(\ell^*+1)}} \alpha_e \var_e   \\
  &\geq \sum_{e \in M^{(\ell^*+1)}} \alpha_{e_{\ell^*+1}} \var_e  \nonumber \\ 
  & = \alpha_{e_{\ell^*+1}} \sum_{e \in M^{(\ell^*+1)}}  \var_e    > \alpha_{e_{\ell^*+1}}  B. \label{eqn_ml2}
\end{align}

Now we show upper bounds on the optimum solution to the BR-MWM problem $M^{OPT}$.  We divide $M^{OPT}$ into two parts: $M^{OPT}_1$ and $M^{OPT}_2$, where the first part is the set of edges in $M^{OPT} \cap H^{(\ell^*)}$ and the second part is the edges not present in $H^{(\ell^*)}$. We present separate upper bounds on $M^{OPT}_1$ and $M^{OPT}_2$. By definition, $M^{OPT}_1$ is a matching on the set of edges $H^{(\ell^*)}$. Therefore, its reward is smaller than or equal to that of the optimum matching on $H^{(\ell^*)}$, which happens to be $M^{(\ell^*)}$. Hence,  
\begin{equation}
\label{eqn_ub1}
R(M^{OPT}_1) \leq R(M^{(\ell^*)}).
\end{equation}

Next, consider $M^{OPT}_2$.  To upper-bound   $R(M^{OPT}_2)$ we also use inequalities~\ref{eqn_ml1},\ref{eqn_ml2}: 
\begin{align*}
R(M^{OPT}_2) & = \sum_{e \in M^{OPT}_2}r_e = \sum_{e \in M^{OPT}_2} \alpha_e \var_e  \\ 
&\leq  \sum_{e \in M^{OPT}_2}\alpha_{e_{\ell^*+1}} \var_e  = \alpha_{e_{\ell^*+1}} \sum_{e \in M^{OPT}_2} \var_e \\ 
&\leq \alpha_{e_{\ell^*+1}} B < R(M^{(\ell^*+1)}) \\  
&\leq 	R(M^{(\ell^*)}) +  r(e_{\ell^*+1}). \label{eqn_ub2}
\end{align*}

\noindent  Now, we are ready to complete the proof. Recall that the output of the algorithm $M_{OUT} $ satisfies $R(M_{OUT}) = \max(R(M^{(\ell^*)}), r_{e_{\ell^*+1}})$.  Combining the upper bounds for $M^{OPT}_1$ and $M^{OPT}_2$, yields
\begin{align*}
R(M^{OPT}) &\leq R(M^{(\ell)}) + R(M^{(\ell)}) +  r(e_{\ell+1}) \\ 
 & = 2 R(M^{(\ell)}) +  r(e_{\ell+1})  \leq 3R(M_{OUT}).
\end{align*}

\noindent This completes the proof. 
\end{proof}

{\em Running time}: Assuming that the $O(mn + n^2\log n)$~\cite{gabow90} implementation of Edmond's algorithm is used as a black-box, we remark that the run time of Algorithm~\ref{pseudo_alg_matchings} is $O(mn\log m + n^2\log m \log n)$.

\spara{Fast $\frac{1}{5}$-approximation} \spara{ for uncertain weighted graphs.}  Since the running time using Edmond's algorithm is prohibitively expensive, we show how the approximation guarantee changes when  we use the (much faster) greedy algorithm for maximum weighted matchings as {\sc Match-Alg}. Recall, the greedy matching algorithm runs in $O(m\log m+n)$ time.

\begin{theorem}
If the black-box {\sc Match-Alg} is set to be the greedy matching algorithm, then Algorithm~\ref{pseudo_alg_matchings} computes a $\frac{1}{5}$-approximation to the optimal  solution of the BR-MWM problem in $O(m \log^2m + n \log m)$-time.
\end{theorem}

The proof is omitted as it is essentially identical to the proof of  Theorem~\ref{thm_graphmatchings}, with the only change that the greedy matching algorithm provides a $\frac{1}{2}$-approximation to the maximum weighted matching problem.

\spara{Fast $\frac{c}{2+c}$-approximation for uncertain weighted hypergraphs.} Recall that finding a maximum weight hypergraph matching  is NP-hard even for unweighted, 3-regular hypergraphs \cite{karp1972reducibility}. However, there exist various  algorithms, that achieve different approximation factors $c <1$. For example, the greedy algorithm provides a $\frac{1}{k}$ approximation guarantee, where $k$ is the rank of the hypergraph (i.e., any hyperedge contains at most $k$ nodes).  Our main theoretical result follows.

\begin{theorem}
\label{thm:hyper}
Given any  $c$-approximation, polynomial-time algorithm {\sc Match-Alg} ($c \leq 1$) for the maximum weighted hypergraph matching problem, we can compute in polynomial time a hypermatching $M_{OUT}$ such that its risk is at most $B$ and its expected weight is  a $\frac{c}{2+c}$-approximation to the expected weight of the optimal hypermatching that has risk  at most $B$.
\end{theorem}

\noindent Again the proof proceeds step by step as the proof of Theorem~\ref{thm_graphmatchings}, and is omitted. In what follows, we restrict our attention to using the greedy hypermatching algorithm as a black-box. Our focus on greedy matchings stems from the fact that its approximation factor ($\frac{1}{k}$) is asymptotically optimal~\cite{berman2000d,chan2012linear}, that it is easy to implement, and runs in $O(m \log m+n)$ time using appropriate data structures. Since we will be using the greedy algorithm in our experiments (Section~\ref{sec:exp}), we provide the following corollary.  

\begin{corollary}
For any hypergraph of rank $k$, we can compute in poly-time a hypergraph-matching whose risk is at most $B$ and whose weight is a $\Omega(\frac{1}{k})$ approximation to the optimum bounded-risk hypergraph  matching.
\end{corollary}

Algorithm~\ref{pseudo_alg_matchings} using the greedy hypermatching algorithm in lieu of {\sc Match-Alg} runs in $O(m\log^2 m+n \log m)$ time.

\spara{Remark.} We reiterate the point that our algorithm can be used to compute risk-averse matchings for other notions of risk such as variance. For instance, if we define risk as in Equation~(\ref{eqn_risk_variance}), then the only thing that changes in our algorithm is the definition of the $\alpha_e$, namely that $\alpha_e$ is set equal to $\frac{r_e}{\var_e^2}$ for each (hyper)edge $e \in E$. The rest, including the theoretical guarantees remain identical.  

%% file: Experiments.tex
\subsection{Experimental Setup and Normalization} 

 We test our proposed algorithm on a diverse range of datasets, where the orders of magnitude of risk (e.g., standard deviation) can vary greatly across datasets. In order to have a consistent interpretation of the trade-off between  expected reward and risk across datasets, we normalize the allowed risk $B$ relative to the maximum possible standard deviation of a benchmark matching, $B_{\max}$. For the purpose of computing or more precisely approximating $B_{\max}$, we run the greedy matching algorithm on the (hyper)graph $G$ ($H$) where the weight on edge $e$ is $\sigma_e$, and set $B_{\max}$ to be the aggregate risk of the computed matching.  While in theory one may observe a matching with greater risk than the obtained value  $B_{\max}$, this does not occur in any of our simulations. We range $B$ according to the rule:

$$ B = B_n \times B_{\max},$$

\noindent where $B_n \in [0,1]$ and is incremented in steps of $0.05$. We refer to $B_n$ as the {\em normalized risk} from now on. 

\spara{Code.} We implement our proposed fast approximation algorithm for uncertain weighted hypergraphs in Python.  The code is available at Github \cite{tsourolampis}.

\spara{Machine specs.} All experiments were performed on a laptop with 1.7 GHz Intel Core i7 processor and 8GB of main memory.

\subsection{Controlled Experiments}
\label{sec:synth}

\spara{Synthetic experiments.} We experiment with two random graph topologies \cite{frieze2015introduction}: Erd\H{o}s-R\'enyi random graphs $G(n,p)$,  and preferential attachment graphs, generated according to the Barab\'asi-Albert $BA(n,m)$ model.  For each normalized risk bound $B_n$, we generate $4$ random graphs that in expectation have  90\,000  edges. For $G(n,p)$ we set $n=6\,000$, and $p=0.005$. The resulting graphs are connected, as $p$ is above the connectivity threshold  $\frac{\log n}{n}$.    For the Barab\'asi-Albert model we set $n=6\,000$, $m=15$. For each random graph we generate, we  choose the weights according to some distribution. Once we have fixed the weights we sample edge probabilities according to some probability distribution. This procedure generates uncertain weighted Bernoulli graphs.     In a similar way we  generate uncertain weighted Gaussian graphs, by first sampling means, and then the variances.    

Specifically, for {\it uncertain weighted Bernoulli graphs}, we sample weights independently  four times  from  (i) uniform $U(0,1000)$, and  (ii) Gaussian $N(100,\frac{100}{6}^2)$.   Then, for each choice of weights,  we create  four different Bernoulli probability settings. We sample probabilities according to (i) uniform $U(0,1)$, and (ii) Gaussian $\mathcal{N}(0.5,\frac{1}{6}^2)$ distributions.   Notice that for the Gaussian distribution, we carefully set the variance at the same order of magnitude as the mean, to allow a  greater range of values.

For {\it uncertain weighted Gaussian graphs}  we first sample the means independently  from (i) uniform $U(0,1\,000)$, and (ii) Gaussian $N(100,\frac{100}{6}^2)$ distributions.  Again, we sample four times for each distribution.  For each choice of mean, we create four different Bernoulli variance settings. We sample the variance of each edge independently from (i) uniform $U(0,100)$, or Gaussian $\mathcal{N}(50,\frac{50}{6}^2)$ distributions.

\begin{figure*}
	\centering
	\includegraphics[width=\linewidth]{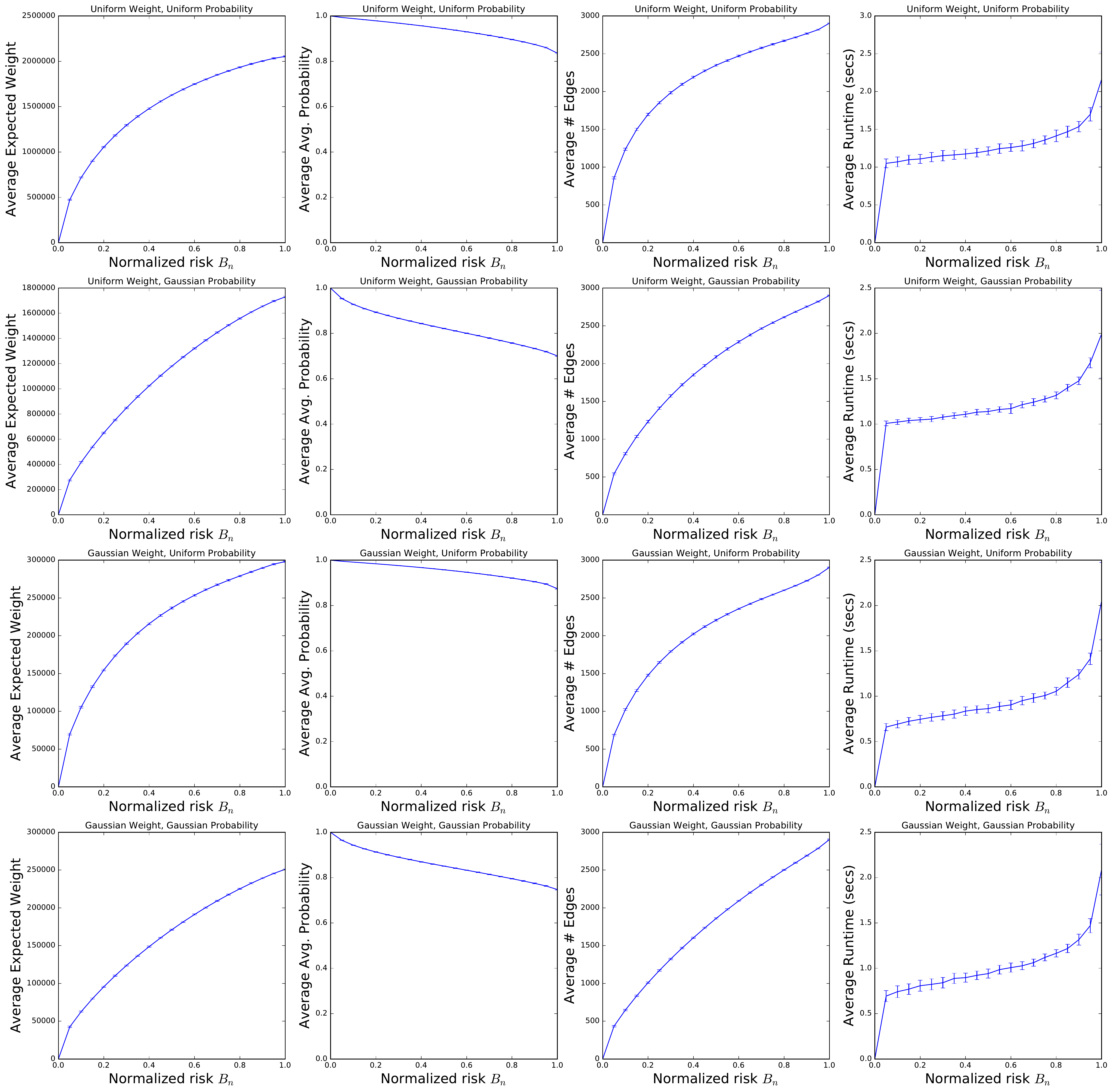} \\ 
	\caption{\label{fig:synth-all} Per column: average (avg.) expected weight, avg. probability, avg. number of edges in the output matching, and the avg. run time of our greedy approximation algorithm vs. the normalized risk bound $B_n$ across different choices of probability distributions for weights, and probabilities on uncertain weighted Bernoulli Erd\H{o}s-R\'enyi graphs. For details, see Section~\ref{sec:synth}.} 
\end{figure*}

\begin{figure*}
	\centering
	\includegraphics[width=\linewidth]{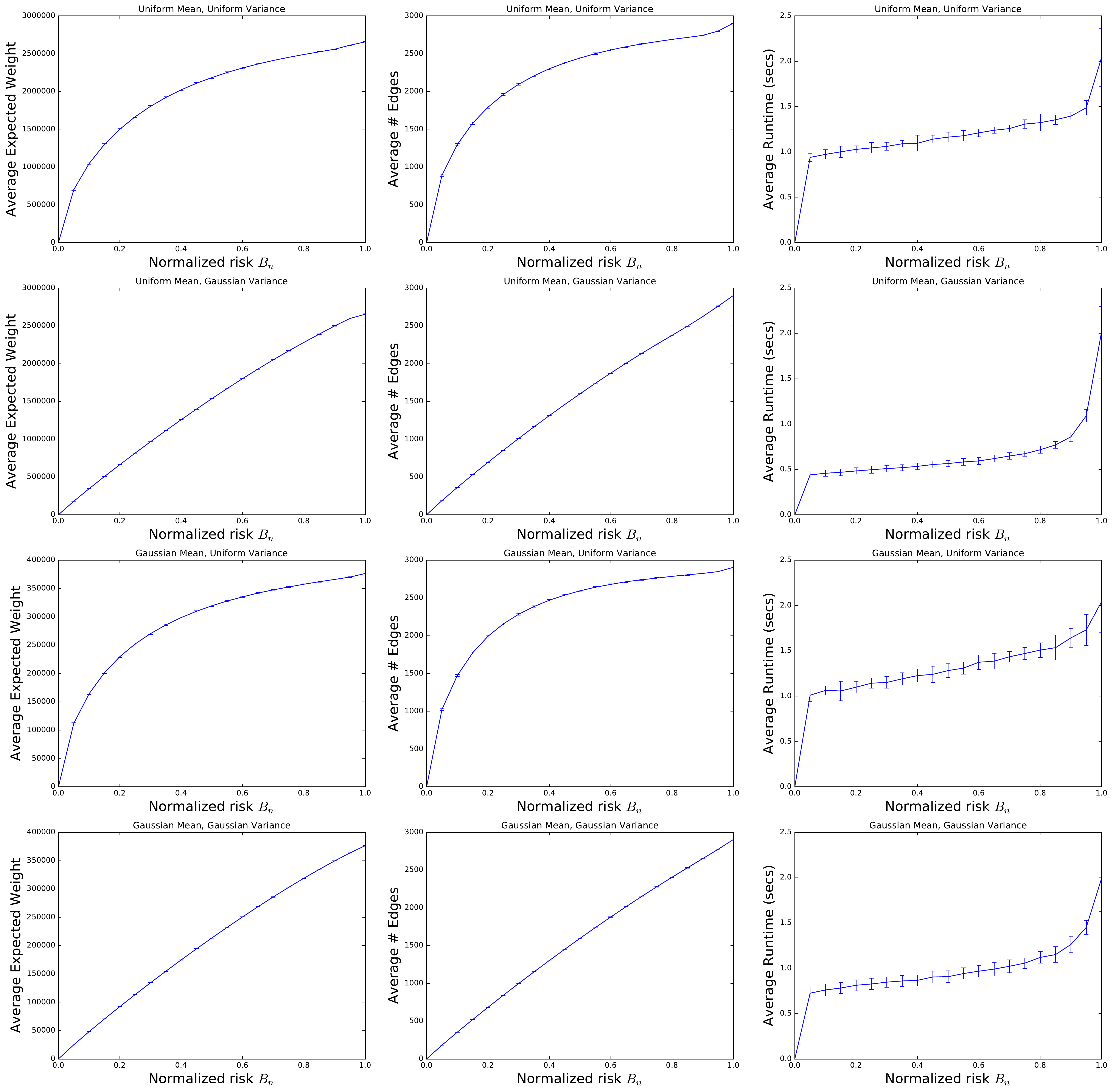} 
	\caption{\label{fig:synth-all2} Per column: average (avg.) expected weight,  avg. number of edges in the output matching, and the avg. run time of our greedy approximation algorithm vs. the normalized risk bound $B_n$ across different choices of probability distributions for weights, and probabilities on uncertain weighted Gaussian Erd\H{o}s-R\'enyi graphs. For details, see Section~\ref{sec:synth}.} 
\end{figure*}

\begin{figure*}
	\centering
	\includegraphics[width=0.9\linewidth]{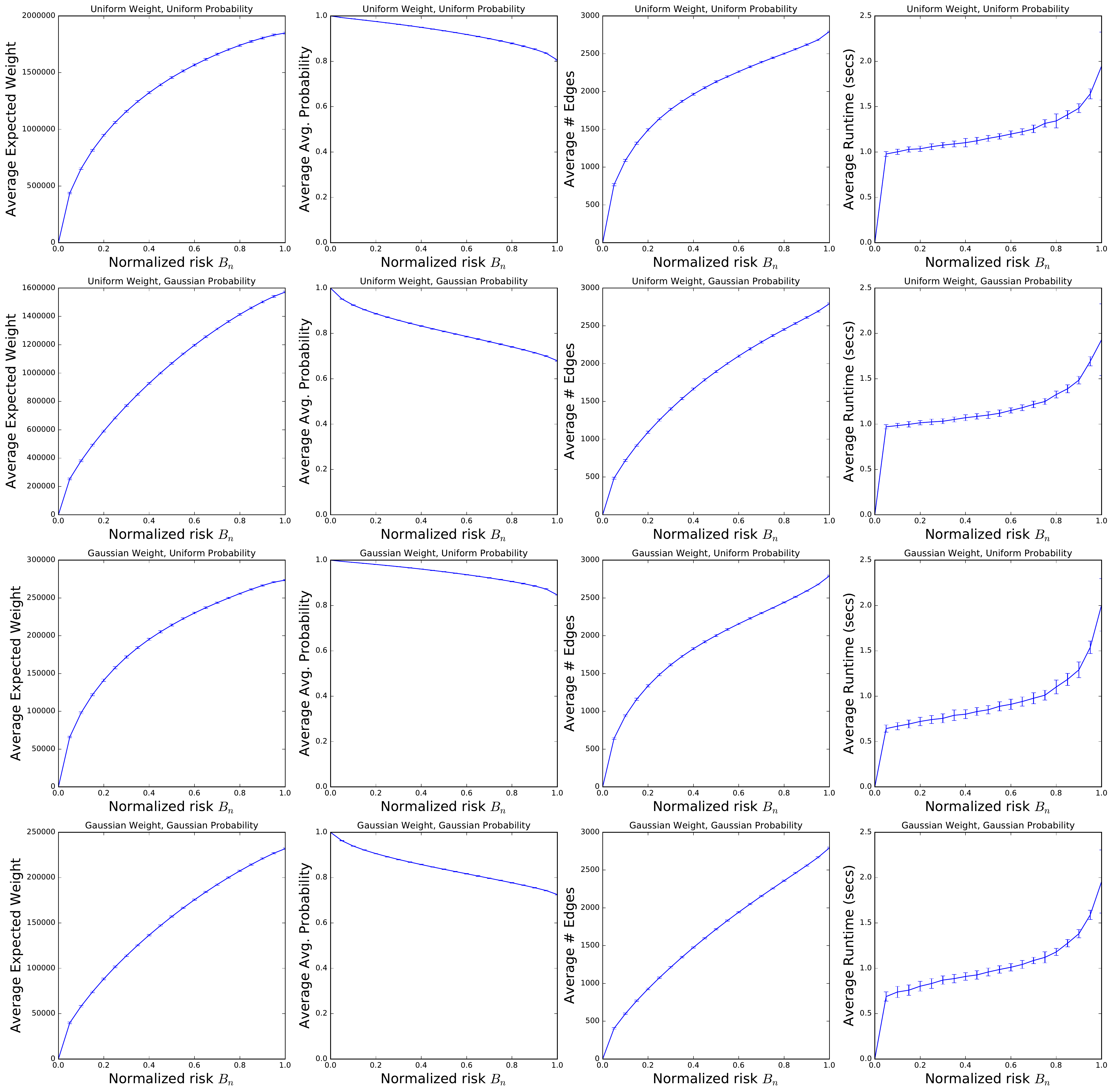} 
	
	\caption{\label{fig:synth-ba1} Per column: average expected weight (avg.), average avg. probability, avg. number of edges in the output matching, and the avg. run time of our greedy approximation algorithm vs. the normalized risk bound $B_n$ across different choices of probability distributions for weights, and probabilities on uncertain weighted Bernoulli Barab\'asi-Albert graphs. For details, see Section~\ref{sec:synth}.} 
\end{figure*}

\begin{figure*}
	\centering
	\includegraphics[width=0.9\linewidth]{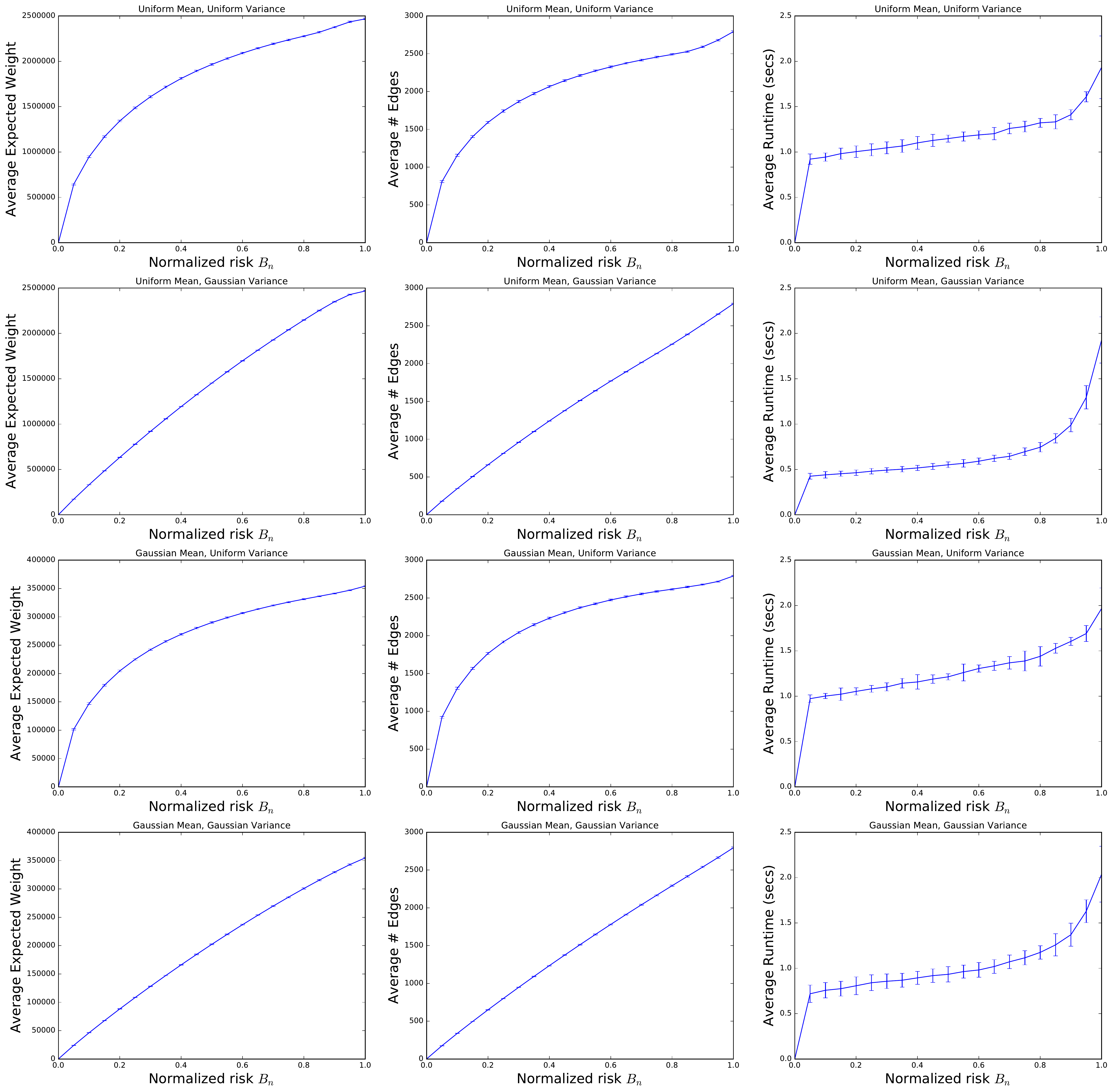} 
	
	\caption{\label{fig:synth-ba2} Per column: average (avg.) expected weight, avg. number of edges in the output matching, and the avg. run time of our greedy approximation algorithm vs. the normalized risk bound $B_n$ across different choices of probability distributions for weights, and probabilities on uncertain weighted Gaussian Barab\'asi-Albert graphs. For details, see Section~\ref{sec:synth}.} 
\end{figure*}

Figure~\ref{fig:synth-all} plots our findings for the case of uncertain weighted Bernoulli graphs on Erd\H{o}s-R\'enyi topologies.  Each plot shows the averages with error bars showing the variability of our findings. Overall, the results tend to be  well concentrated.  In all plots the $x$-axis corresponds to the normalized risk bound $B_n$. 
Each row corresponds to a different setting of sampling distributions for the edge weights, and the edge probabilities.  The first row corresponds to choosing both the weights and the edge probabilities uniformly at random, the second row  to uniform weights, and Gaussian probabilities, the third to Gaussian weights, and uniform probabilities, and the fourth to Gaussian weights, and Gaussian probabilities respectively. The first column corresponds to the average expected weight, i.e., the expected weight of each matching, averaged over all experiments per $B_n$ value, the second column to the average probability of the hyperedges chosen in the matching, averaged  over all experiments,   the third to the average number of edges in the matching, and the fourth to the average run time. We observe similar results across all settings in how the objective changes  
as a function of the normalized bound $B_n$.

Figure~\ref{fig:synth-all2} shows our findings for uncertain weighted Gaussian graphs using an Erd\H{o}s-R\'enyi topology. Since the edge weight distribution is continuous, there is no plot for average probability as in the case of uncertain Bernoulli graphs. We observe that the expected reward is greater for  Erd\H{o}s-R\'enyi topologies. Interestingly, when the variance is sampled from a Gaussian, the growth of number of edges in the output matching is a linear function of $B_n$. A positive side-effect of risk-aversion is faster run times: the smaller the $B_n$, the faster the algorithm completes. 

The corresponding plots for the Barab\'asi-Albert topologies with both Bernoulli and Gaussian distributions are presented in Figures~\ref{fig:synth-ba1} and~\ref{fig:synth-ba2} respectively. These plots are mostly similar to the plots in Figures~\ref{fig:synth-all} and~\ref{fig:synth-all2} and indicate the same kind of trends.   Erd\H{o}s-R\'enyi graphs seem to yield matchings with higher expected reward on average when compared to Barab\'asi-Albert graphs.  For large $B_n$ values (e.g., $B_n = 1$), this is to be expected since there is a {\em perfect} matching with high probability in the former graphs for $p$ above the connectivity threshold \cite{frieze1988matchings}. Proving that this relation holds for intermediate $B_n$ values as well is an interesting question.

\begin{figure*} 
	\centering
	\includegraphics[width=.65\textwidth]{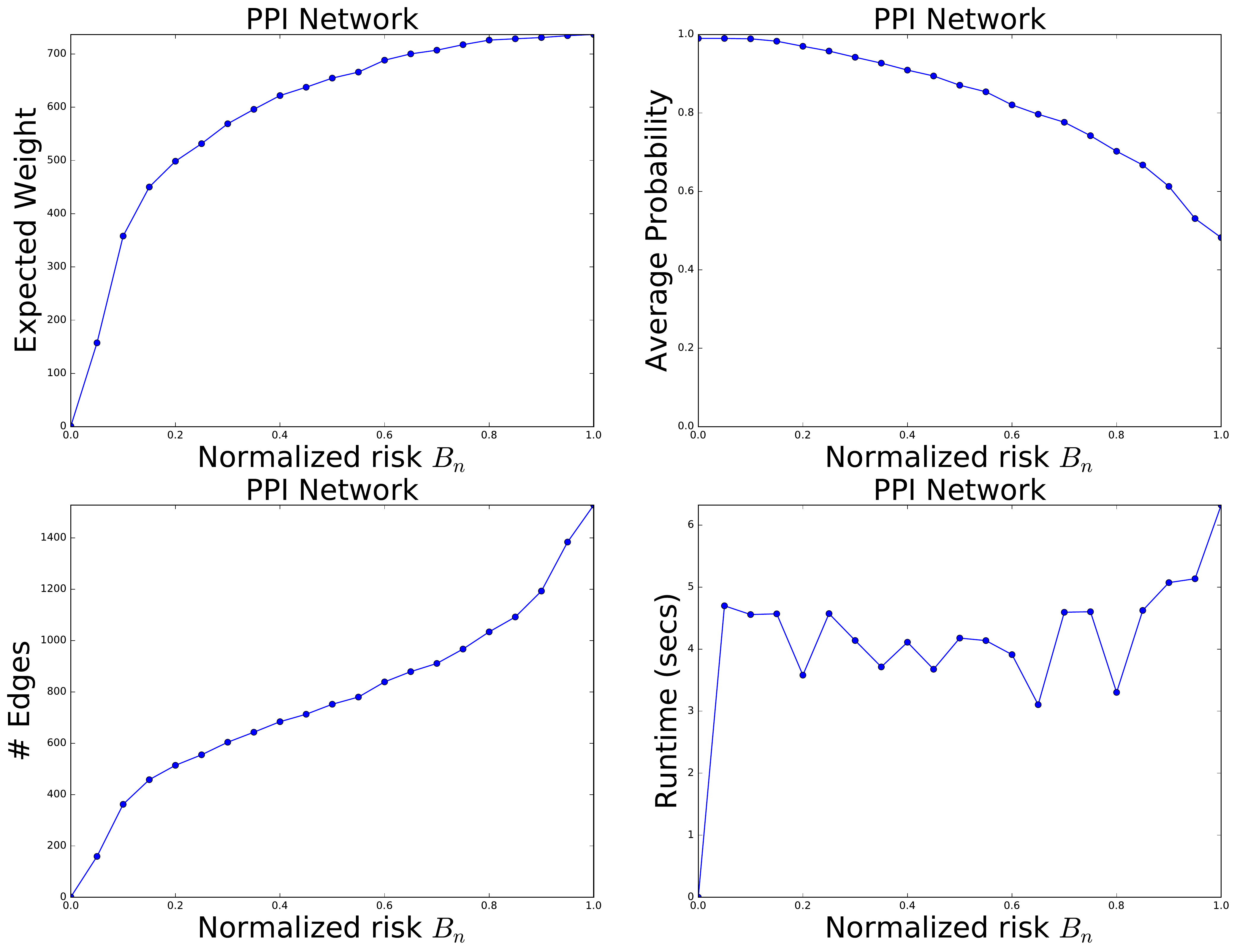}
	\caption{\label{fig:ppi} (a) Expected reward, (b) average probability (over matching's edges),  (c) number of edges in the matching, and (d) running time in seconds versus normalized risk $B_n$ for the uncertain PPI network. For details, see Section~\ref{sec:synth}.} 
\end{figure*}

 \spara{Uncertain Unweighted PPI network.}  We use a real-world uncertain protein-protein interaction (PPI) network that contains 7\,123 protein-protein interactions involving 2\,708 proteins \cite{krogan2006global}. The input graph is unweighted, i.e., all weights are equal to one. The dataset is publicly available as supplementary material to \cite{krogan2006global}. Figure~\ref{fig:ppi} shows our findings for the PPI network. The observed trends are similar to those seen in the case of synthetic topologies. It is worth noting that when $B_n$ is small, the algorithm quickly picks the most certain edges, and then keeps adding edges with lower probability. 

\subsection{Recommending impactful but probable collaborations} 
\label{sec:dblp}

\spara{Dataset.}  In many ways, academic collaboration is an ideal playground to explore the effect of risk-averse team formation for research projects as there exist teams of researchers that  have the potential for high impact but may also collaborate less often. To explore this further, we use our proposed algorithm for uncertain weighted hypergraphs as a tool for identifying a set of disjoint collaborations that are both impactful and likely to take place. For this purpose, we use the Digital Bibliography and Library Project (DBLP) database.  From each paper, we obtain a team that corresponds to the set of authors of that paper. As a proxy for the impact of the paper we use the citation count. Unfortunately, we could not obtain  the citation counts from Google Scholar for the whole DBLP dataset as we would get rate limited by Google after  making too many requests. Therefore, we used the AMiner  citation network dataset \cite{aminer} that contains citation counts, but unfortunately is not as up-to-date as Google Scholar is. 

We preprocessed the dataset by removing all single-author papers since the corresponding hyperedge probabilities are one. Furthermore, multiple hyperedges are treated as one, with citation count equal to the sum of the citation counts of the multiple hyperedges. To give an example, if there exist three papers  in the dataset that have been  co-authored by authors $A_1,A_2$ with citation counts $w_1,w_2,w_3$ we create one hyperedge on the nodes that correspond to  $A_1,A_2$ with weight equal $w_1+w_2+w_3$. If there exists another paper co-authored by $A_1,A_2,A_3$, this yields a different hyperedge/team $\{A_1,A_2,A_3\}$, and we do not include its citations in the impact of  team $\{A_1,A_2\}$. 

For hyperedge $e = (u_1,\ldots, u_{\ell} )$ we find the set of papers $\{P_1,\ldots,P_{\ell}\}$ authored by authors $u_1,\ldots,u_{\ell}$ respectively. We set the probability of hyperedge $e$ as 

$$p_{e} = \frac{|P_{1} \cap P_{2} \cap \ldots \cap P_{\ell}|}{|P_{1} \cup P_{2} \cup \ldots \cup P_{\ell}|}.$$

\begin{figure} 
\centering
\begin{tabular}{cc}
\includegraphics[width=.45\textwidth]{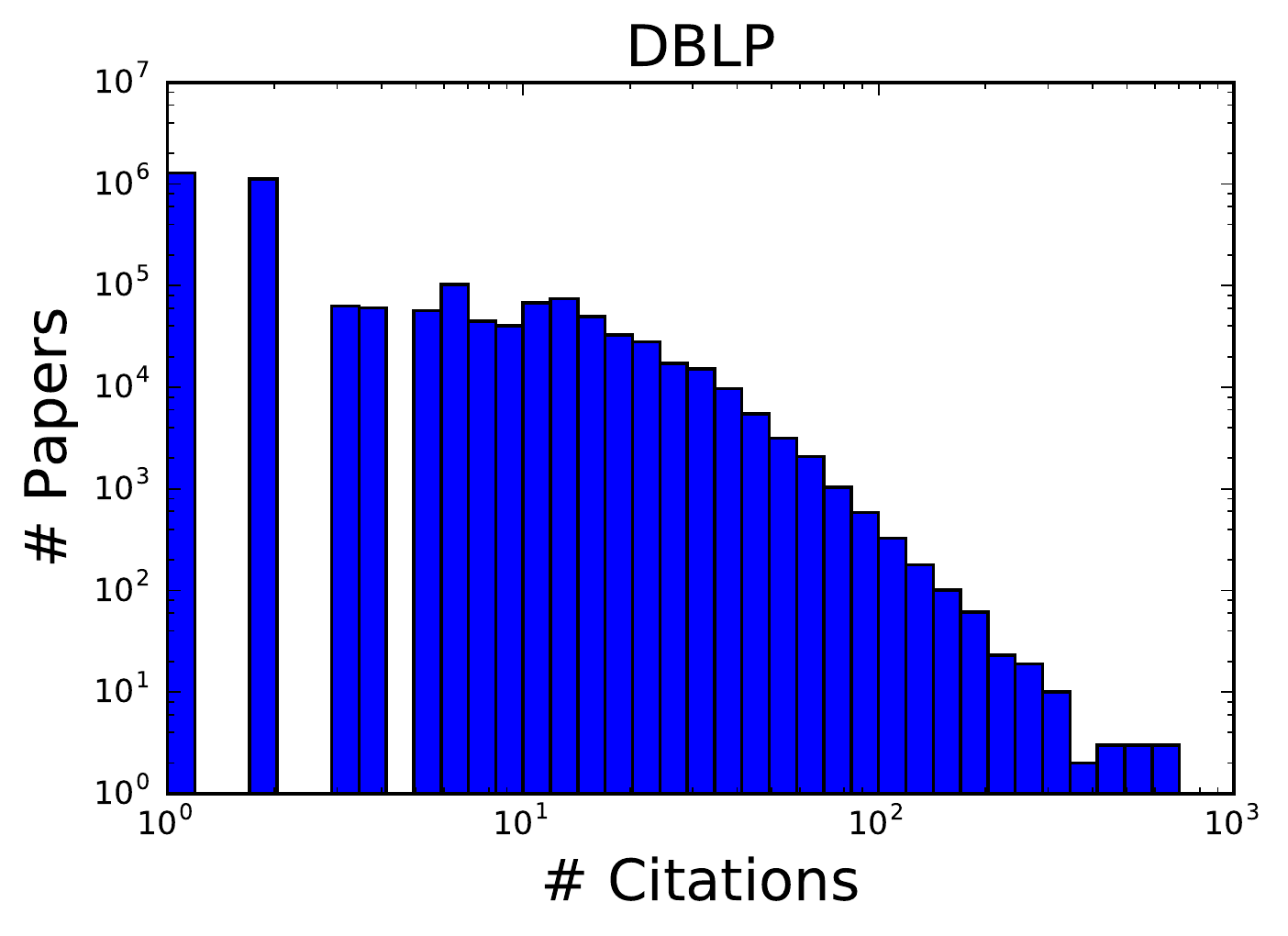} 
	& \includegraphics[width=.45\textwidth]{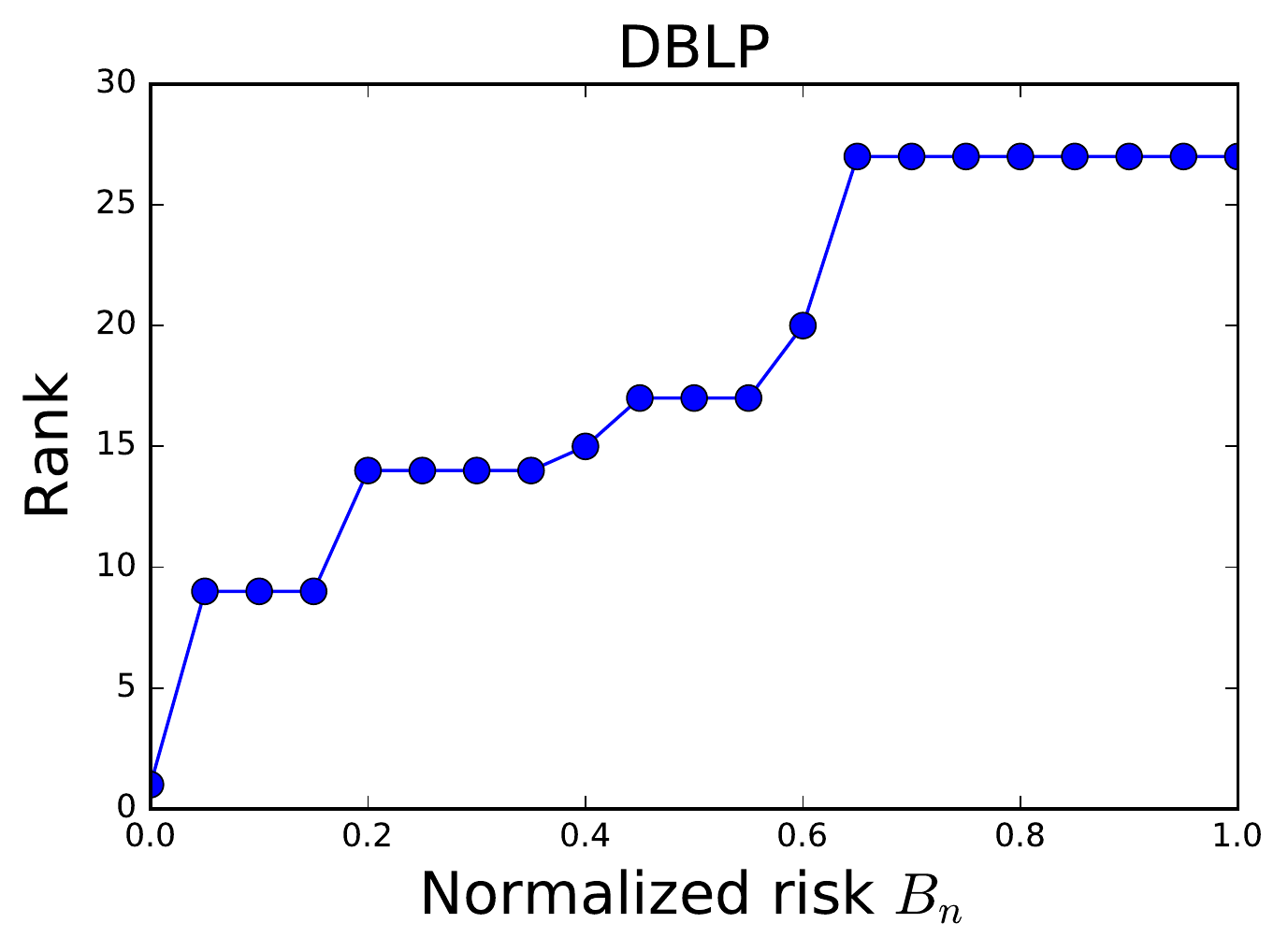}  \\
(a) & (b) 
\end{tabular}
\caption{\label{fig:hyperrank} (a) DBLP citation histogram. (b) Hypergraph rank $k$ versus normalized risk $B_n$. For details, see Section~\ref{sec:dblp}. } 
\end{figure}

\noindent Intuitively, this is the empirical probability of collaboration between the specific set of authors.  

To sum up, we create an uncertain weighted hypergraph using the DBLP dataset, where each node corresponds to an author, each  hyperedge represents a paper whose reward follows a Bernoulli distribution with weight equal to the number of its citations, and probability $p_e$ is the likelihood of collaboration.  The final hypergraph consists of  $n=1,752,443$ nodes and $m=3,227,380$  edges, and will be made publicly available on the first author's website. The largest collaboration involves a paper co-authored by 27 people, i.e., the  rank $k$ of the hypergraph is 27.  Figure~\ref{fig:hyperrank}(a) shows the histogram of citations.

\begin{figure*}[!ht]
	\centering
	\begin{tabular}{cc}
		\includegraphics[width=.43\textwidth]{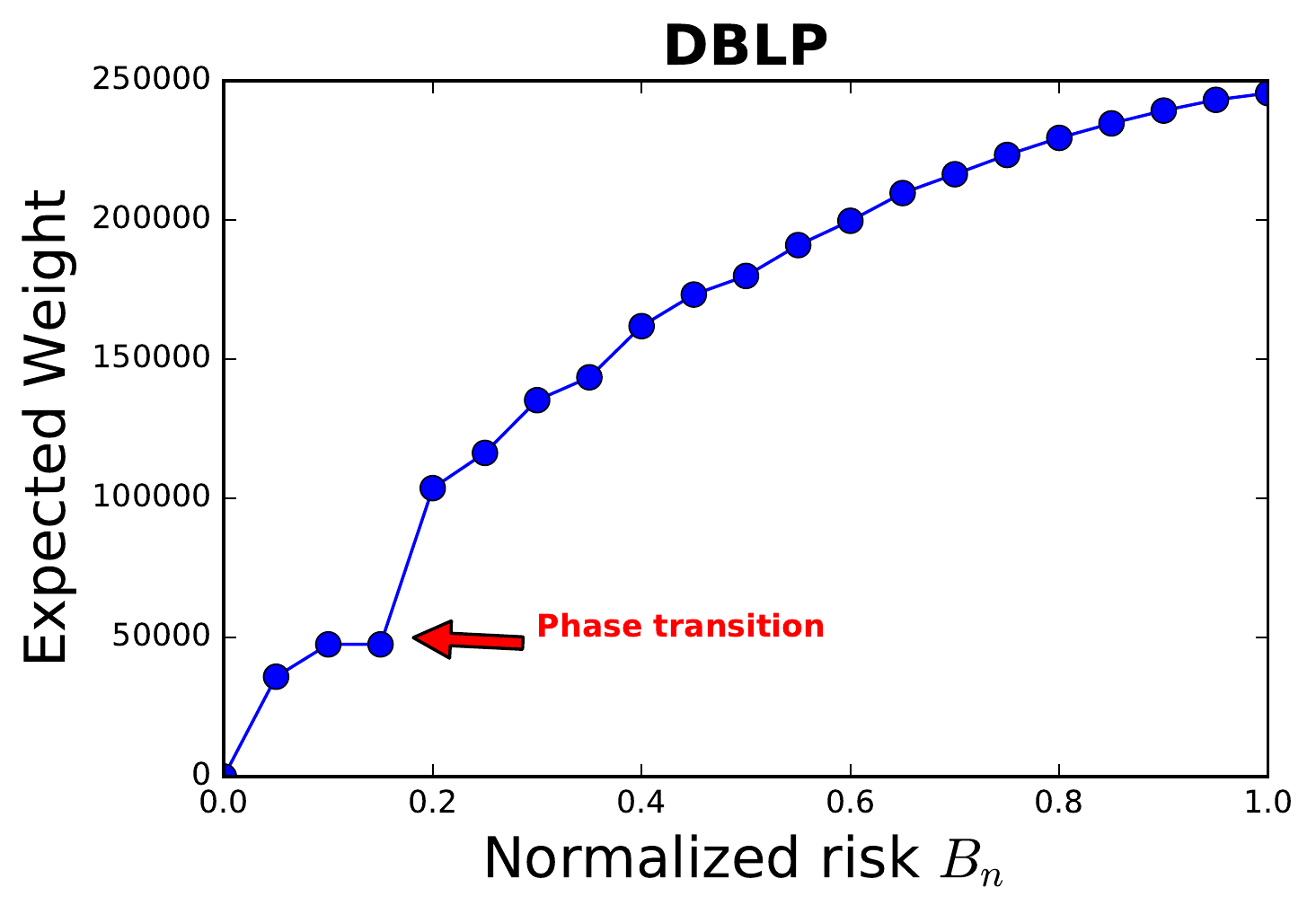}
		& \includegraphics[width=.43\textwidth]{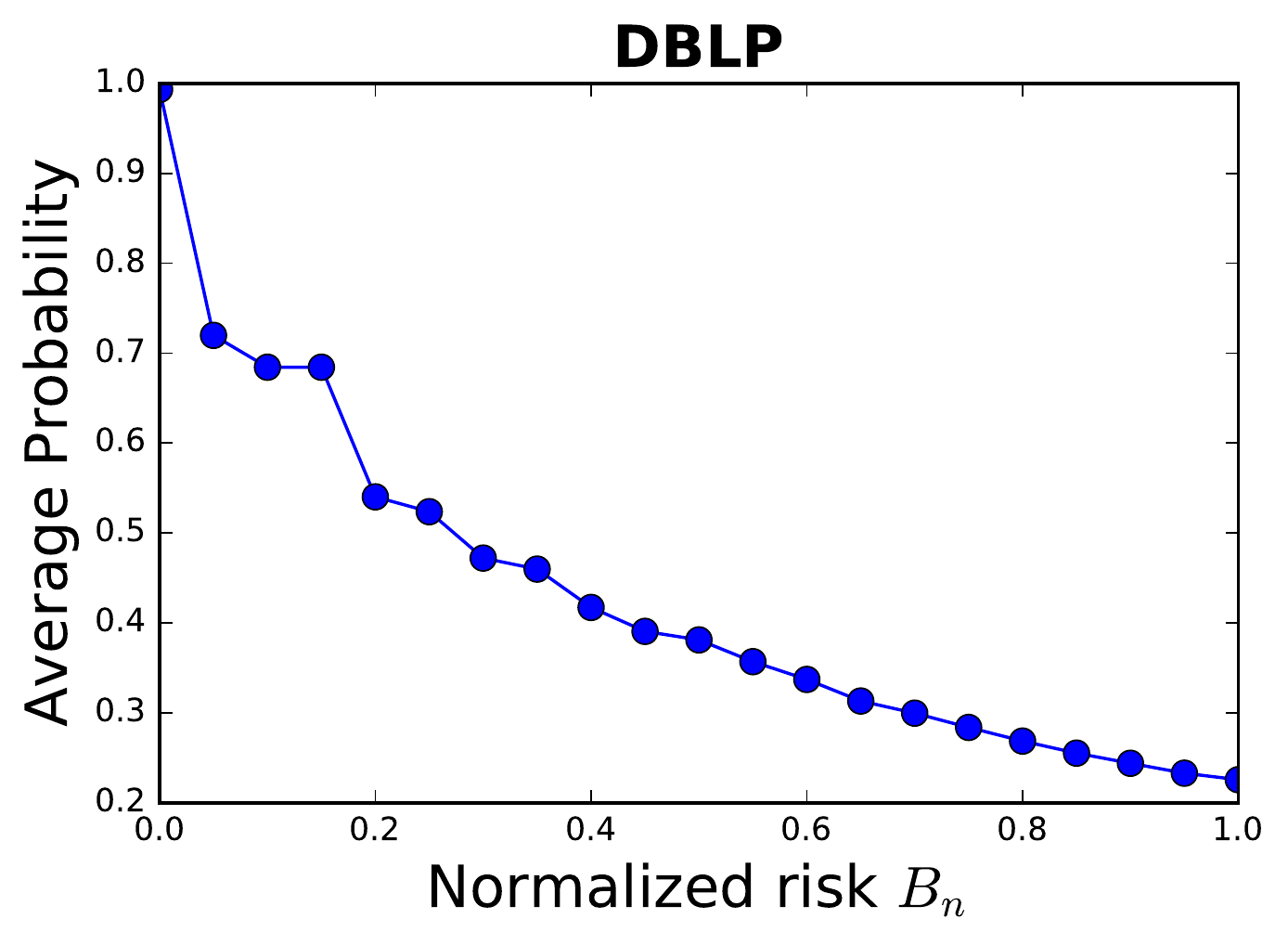} \\ 
		{\small (a)}  &  {\small(b)}  \\ 
		\includegraphics[width=.43\textwidth]{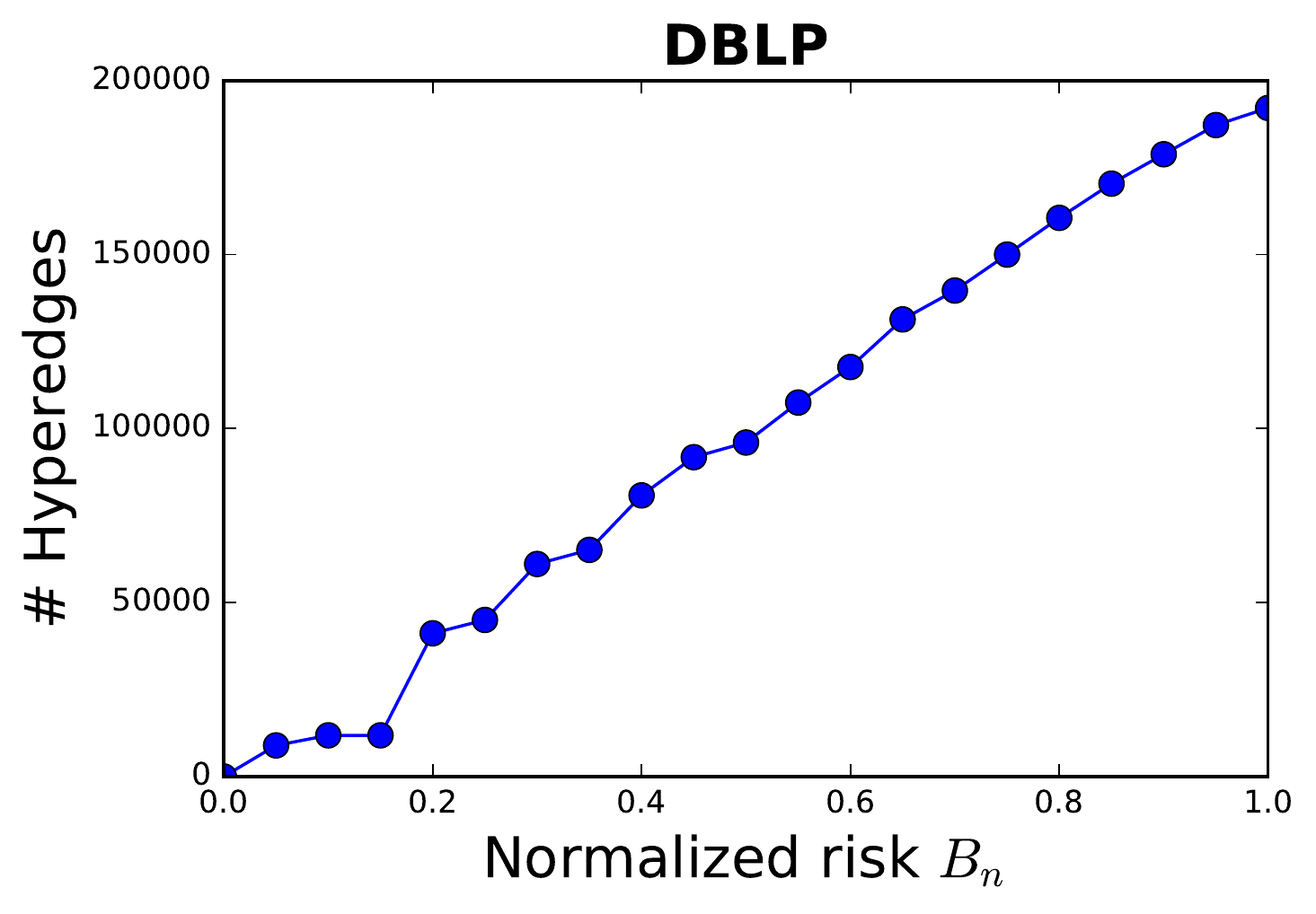} &
		\includegraphics[width=.43\textwidth]{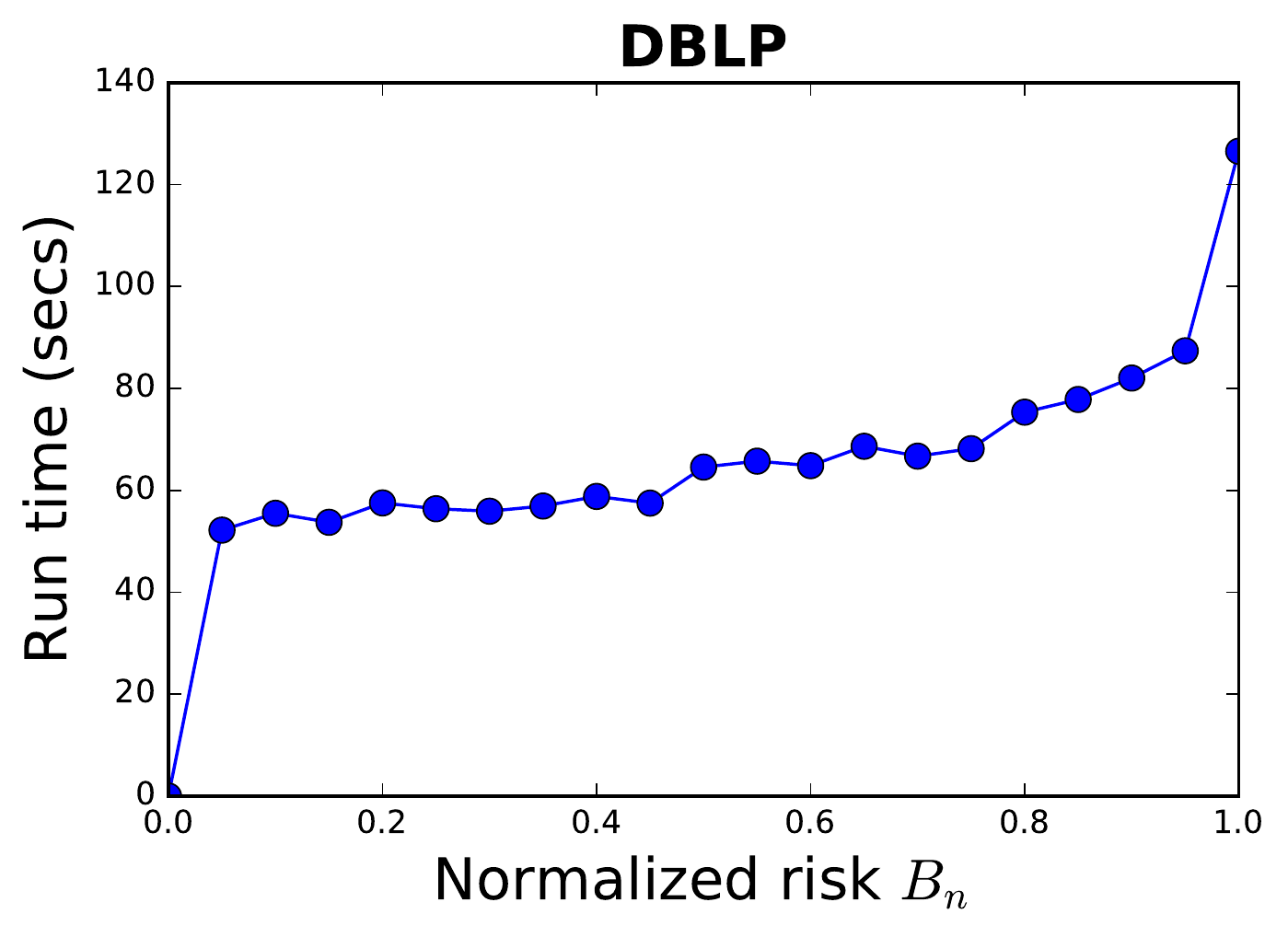}\\
		{\small (c)} &   {\small (d)}\\
	\end{tabular}
	\caption{\label{fig:dblp} (a) Expected reward, (b) average probability (over hypermatching's edges),  (c) number of edges in the hypermatching, and (d) running time in seconds versus normalized risk $B_n$. For details, see Section~\ref{sec:dblp}.} 
\end{figure*} 
\vspace{-2mm}
\spara{Results.} Figure~\ref{fig:dblp} shows our findings when we vary the normalized risk bound $B_n$ and obtain a hypermatching for each value of this parameter, using our algorithm. For the record, when $B_n=1$, then $B =B_{\max}=454\,392.0$. Figure~\ref{fig:dblp}(a) plots the expected weight of the hypermatching versus $B_n$. We observe an interesting phase transition when $B_n$ changes from 0.15 to 0.2. This is because after $B_n=0.15$ the average probability of the hyper-matching drops from $\sim 0.7$ to $\sim 0.5$. This is shown in Figure~\ref{fig:dblp}(b) that plots the average probability of the edges in each hypermatching computed by our algorithm vs. $B_n$.  Figures~\ref{fig:dblp}(a),(b) strongly indicate what we verified by inspecting the output: up to $B_n=0.15$, our algorithm picks teams of co-authors that tend to collaborate frequently.  This finding illustrates that our tool may be used for certain anomaly detection tasks. Figures~\ref{fig:dblp}(c),(d) plot the number of hyperedges returned by our algorithm, and its running time in seconds vs $B_n$. We observe that a positive side-effect of using small risk bounds is speed: for small $B_n$ values, the algorithm computes fewer maximum matchings.

By carefully inspecting the output of our algorithm for different $B_n$ values, we see that at low values, e.g., $B_n = 0.05$, we find hyperedges typically with 50 to 150 citations with probabilities ranging typically  from  0.66 to 1.   When $B_n$ becomes large we find hyper-edges with significantly more citations but with lower probability. For example, for $B_n=0.95$ we find the team  of David Bawden, and Lyn Robinson with weight 934 and probability 0.085.   Additionally, we observe that the rank of the  hypergraph we obtain when our algorithm terminates as a function of $B_n$  increases. This is shown in Figure~\ref{fig:hyperrank}(b). This is intuitive as collaborations with many co-authors are less likely to happen regularly.

\begin{figure*}[!ht]
	\centering
	\begin{tabular}{cccc}
		\includegraphics[width=.23\textwidth]{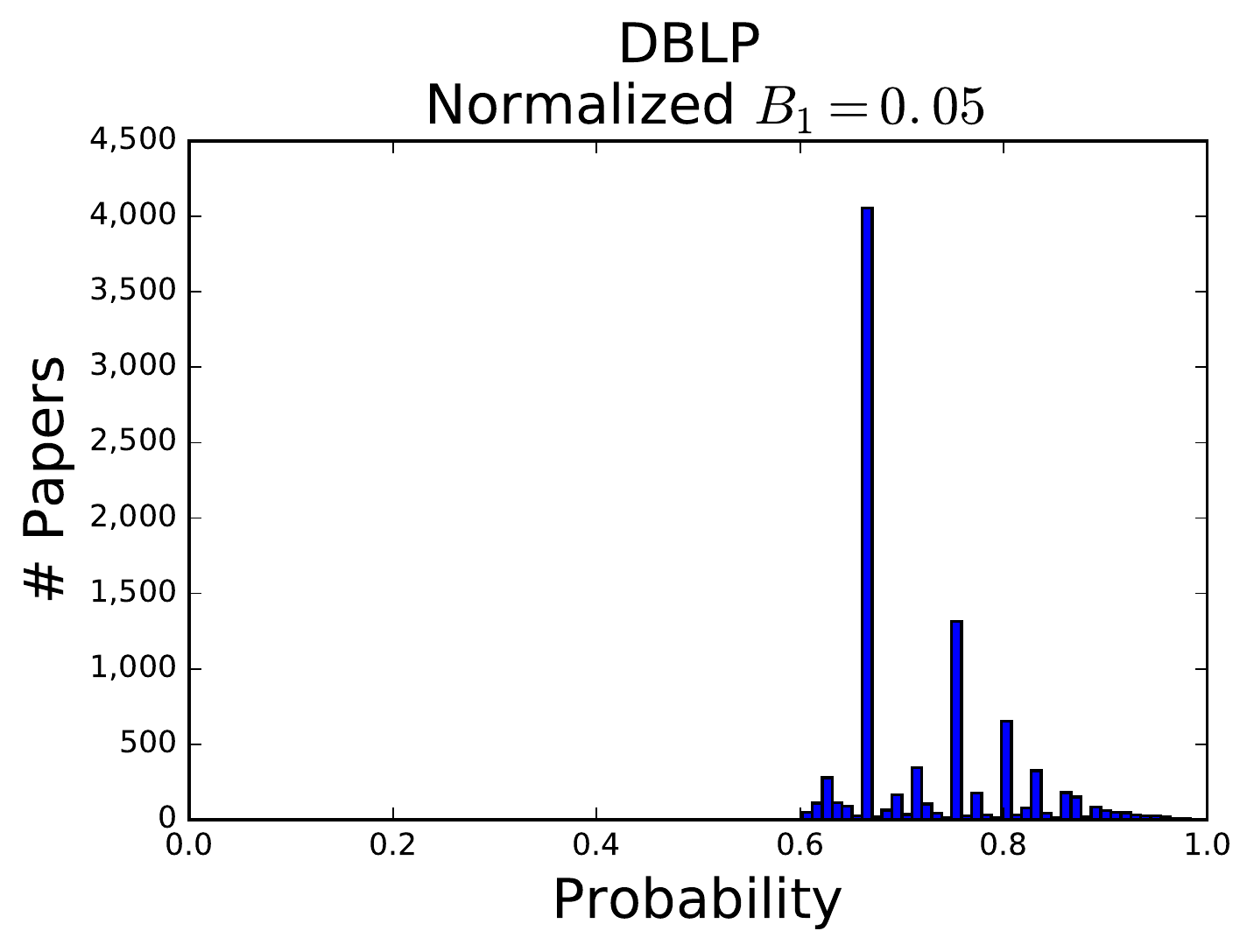}
		& \includegraphics[width=.23\textwidth]{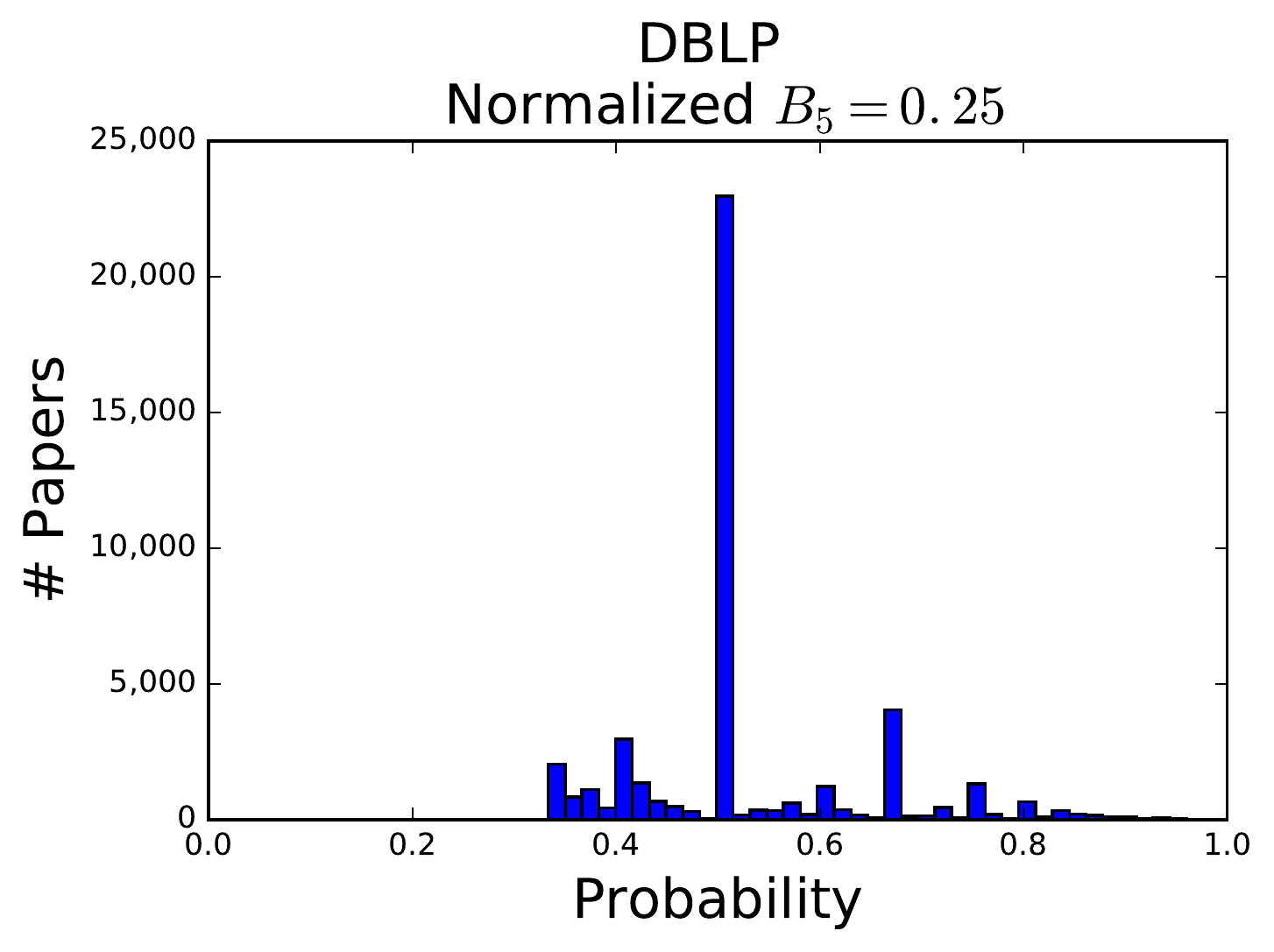}  	& \includegraphics[width=.23\textwidth]{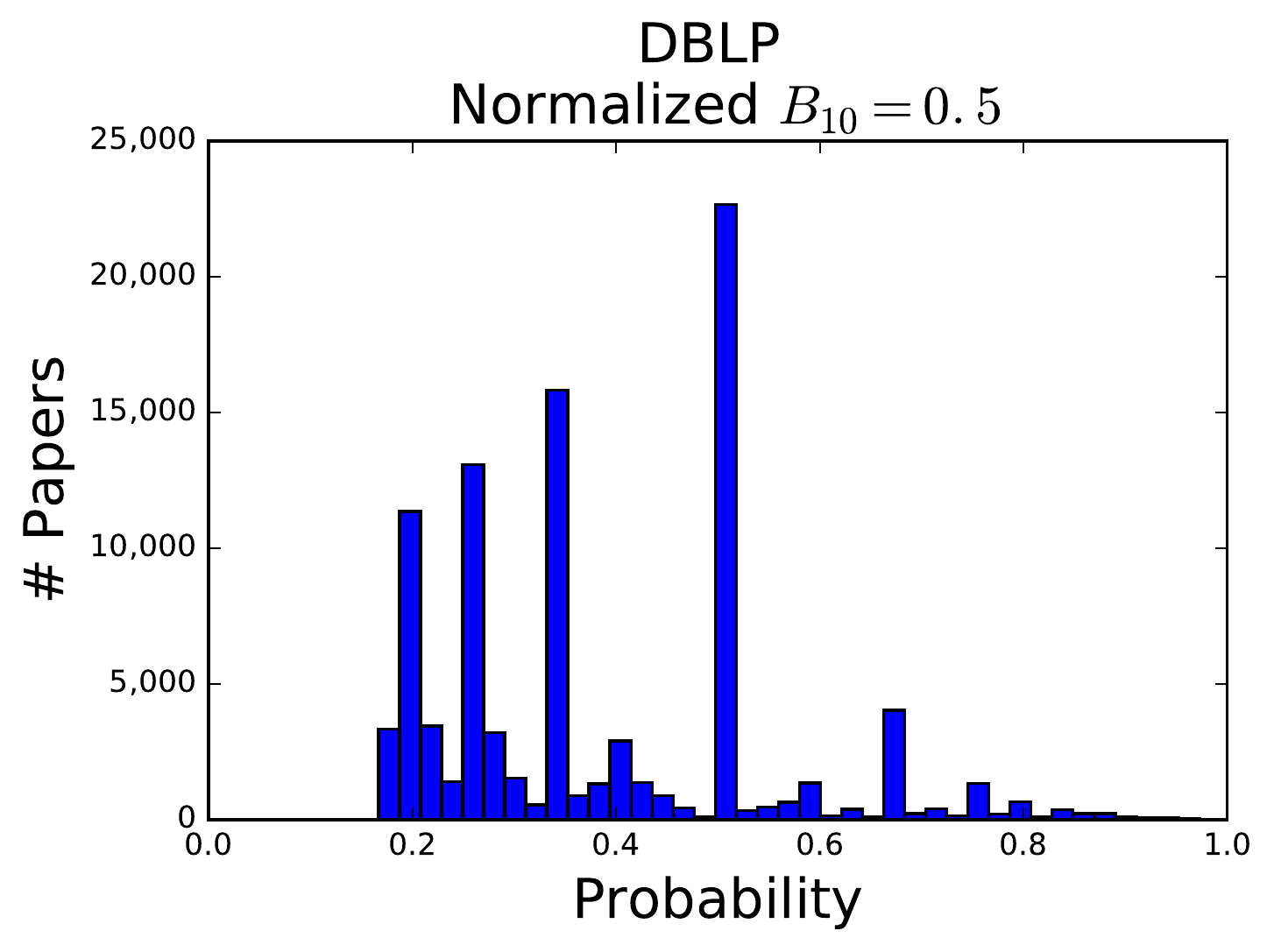}  	& \includegraphics[width=.23\textwidth]{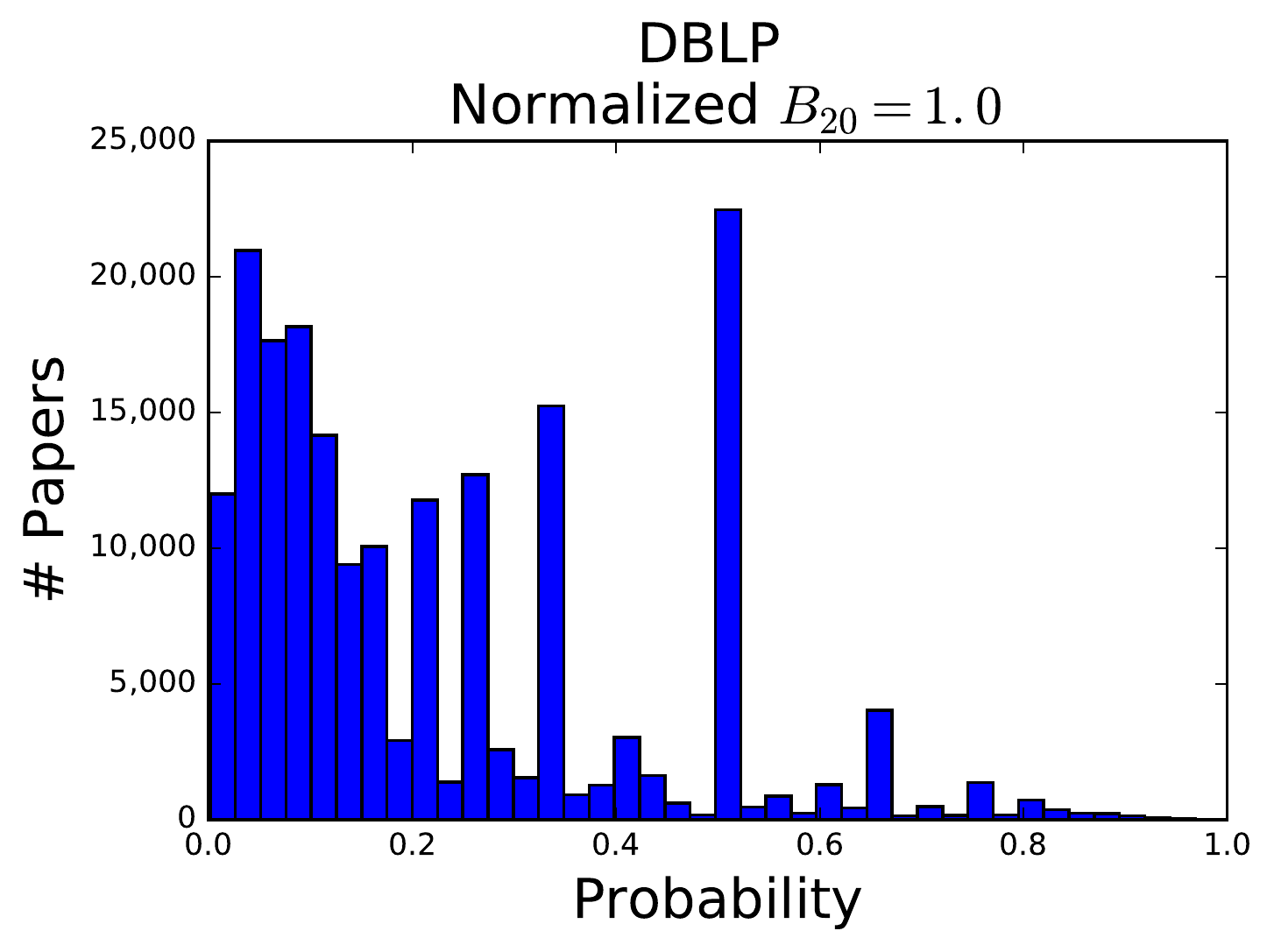}\\   
		{\small ($a_1$)}  &  {\small($b_1$)} & {\small ($c_1$)} &   {\small ($d_1$)} \\ 
			\includegraphics[width=.23\textwidth]{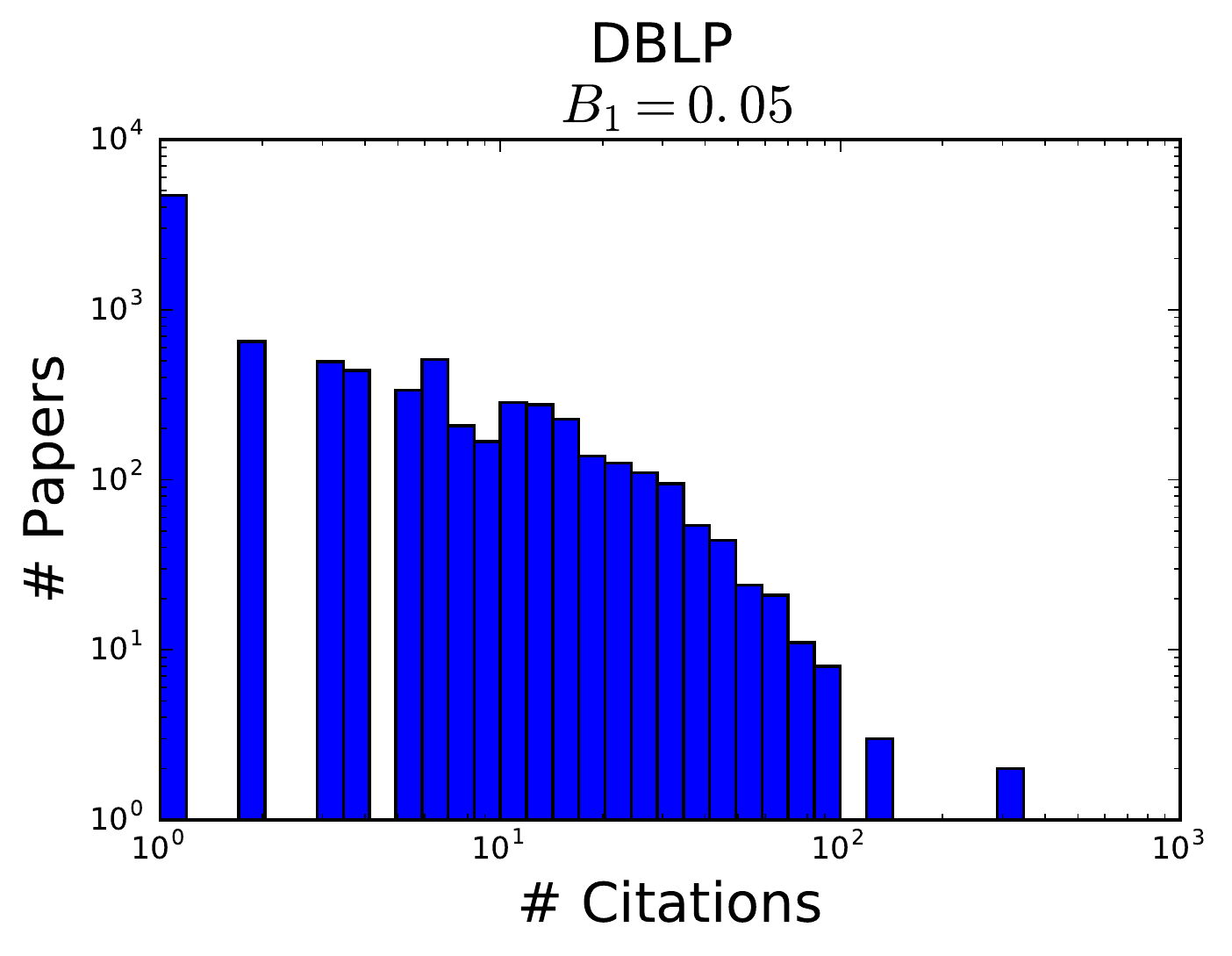}
		& \includegraphics[width=.23\textwidth]{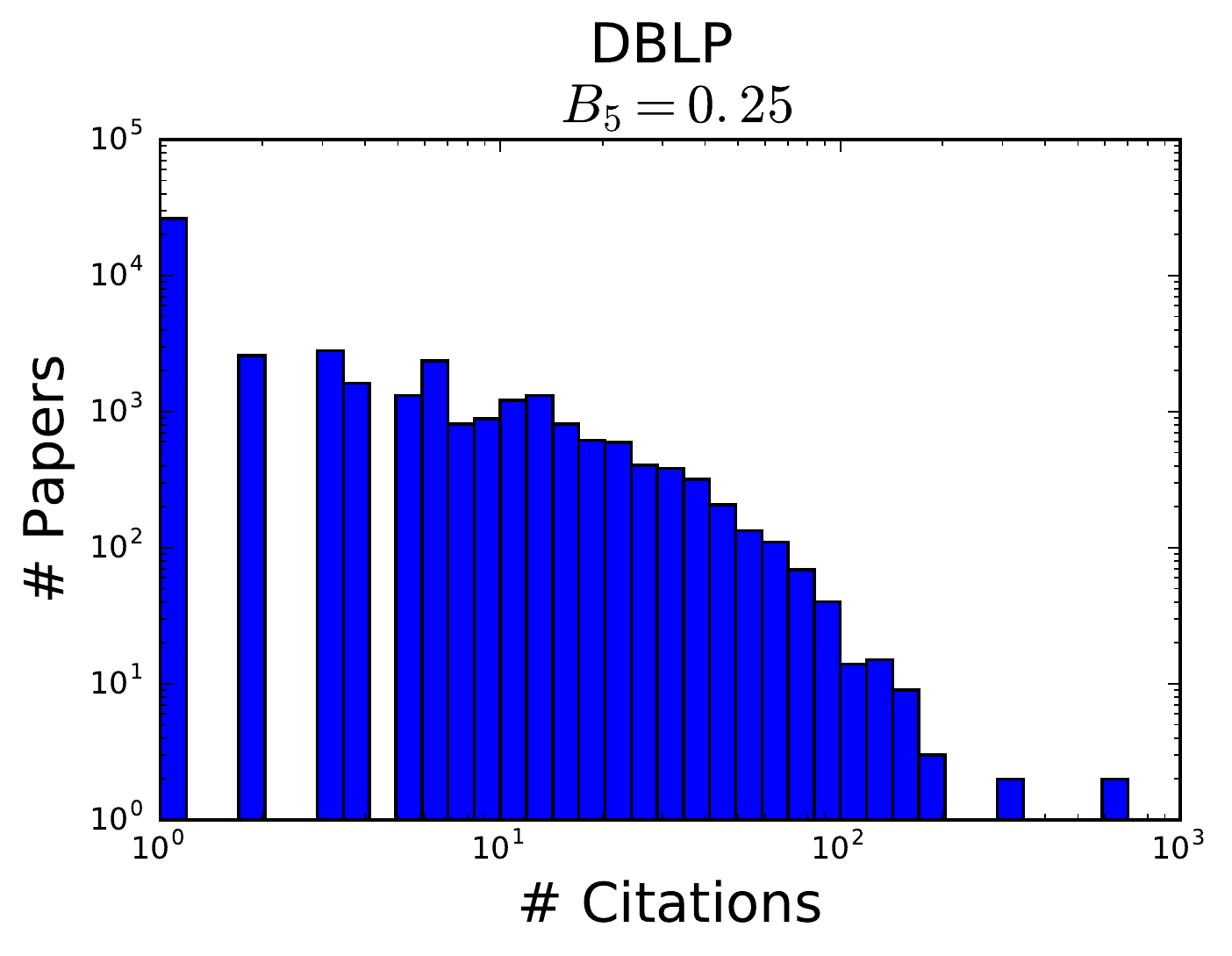}  	& \includegraphics[width=.23\textwidth]{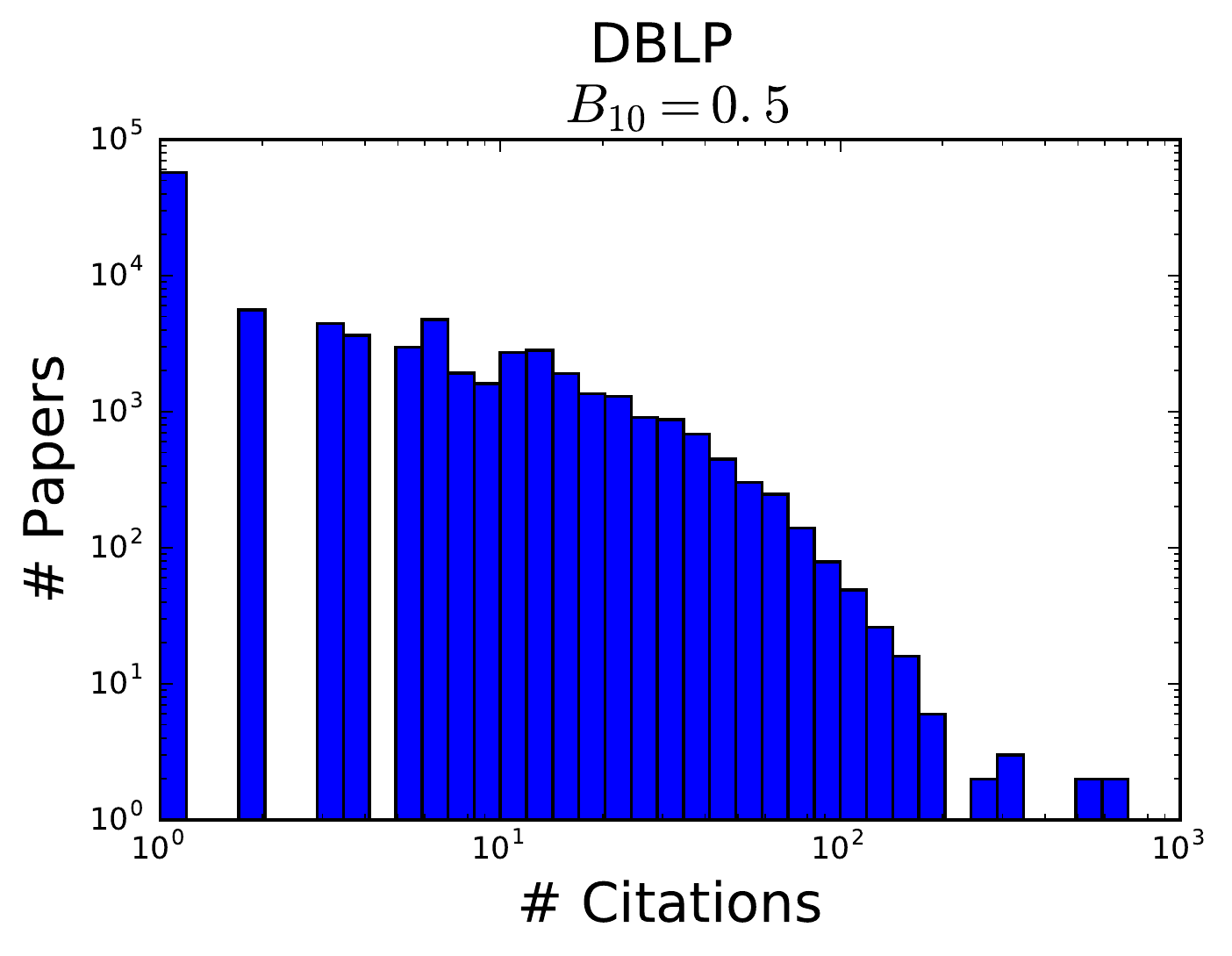}  	& \includegraphics[width=.23\textwidth]{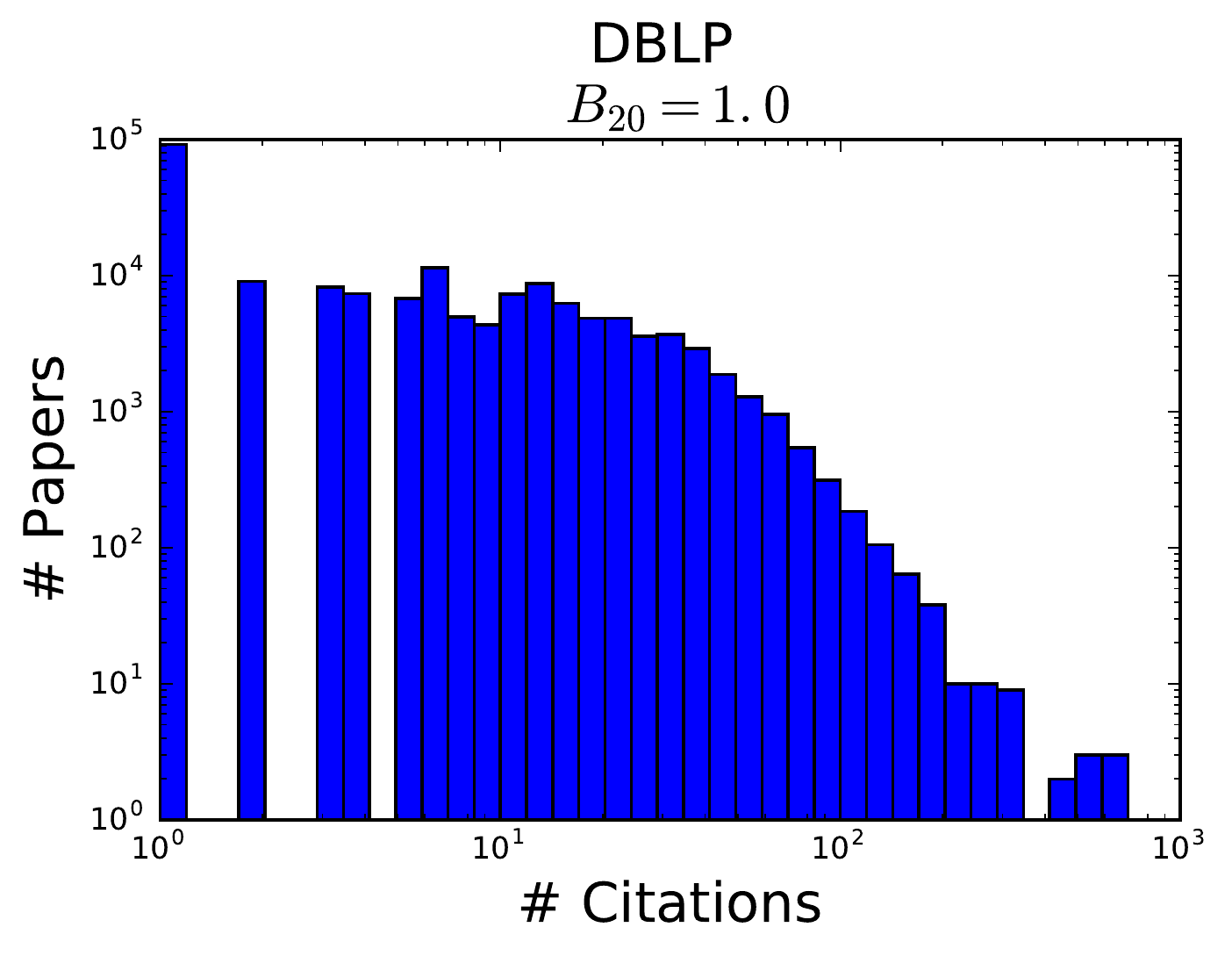}\\   
			{\small ($a_2$)}  &  {\small($b_2$)} & {\small ($c_2$)} &   {\small ($d_2$)} \\ 
	\end{tabular}
	\caption{\label{fig:dblp_citations} Figures in first row $(a_1),(b_1),(c_1),(d_1)$ (second row $(a_2),(b_2),(c_2),(d_2)$): histograms showing the hyperedge probabilities (citations) in the hypermatching returned by our algorithm for  normalized risk values $B_n$  equal to $0.05, 0.25,0.5, 1$ respectively. For details, see Section~\ref{sec:dblp}.} 
\end{figure*} 

Finally,  Figure~\ref{fig:dblp_citations} shows four pairs of histograms 
corresponding to the output of our algorithm for four different normalized risk values $B_n$, i.e., $0.05, 0.25,0.5, 1$ respectively. Each pair ($\{(\small a_1),(\small a_2)\}$, $\{(\small b_1),(\small b_2)\}$, $\{(\small c_1),(\small c_2)\}$, and $\{(\small d_1),(\small d_2)\}$) plots the histogram of the probabilities, and the number of citations of the hyperedges selected by our algorithm for $B_n \in \{ 0.05, 0.25,0.5, 1\}$ respectively. The histograms provide a view of how the probabilities decrease and citations increase as we as we increase  $B_n$, i.e., as we allow higher risk.

%% file: Conclusion.tex
In this work we study the problem of finding matchings with high expected reward and bounded risk on large-scale uncertain hypergraphs. We introduce a general model for uncertain weighted hypergraphs that allows for both continuous and discrete probability distributions, we provide a novel stochastic matching formulation that is NP-hard, and develop fast approximation algorithms. We verify the efficiency of our proposed methods on several synthetic and real-world datasets.
 
In contrast to the majority of prior work on uncertain graph databases, we show that it is possible to combine risk aversion, time efficiency, and theoretical guarantees simultaneously. Moving forward, a natural research direction  is to design risk-averse algorithms for other  graph mining tasks such as motif clustering  \cite{benson2016higher,tsourakakis2017scalable}, the $k$-clique densest subgraph problem \cite{gionis2015dense,tsourakakis2015k}, and $k$-core decompositions \cite{bonchi2014core}?   